\documentclass[11pt,a4paper,oneside]{article}
\usepackage{cmap}
\usepackage[utf8]{inputenc}
\usepackage[top=2cm, bottom=3cm, left=2cm, right=2cm]{geometry}
\usepackage{xspace}
\usepackage{booktabs}
\usepackage{listings}
\usepackage{graphicx}
\usepackage{mathtools,amsmath,amsfonts}
\usepackage{tabulary,tabularx,ragged2e,booktabs,caption}
\usepackage{color}
\usepackage{multirow}
\usepackage{natbib}
\usepackage{amsmath}
\usepackage{hyperref} 
\usepackage[table]{xcolor}
\usepackage[labelfont=bf, font=it]{caption}
\usepackage{float}
\usepackage{imakeidx}
\makeindex

\definecolor{light-gray}{gray}{0.95}

\def\P{\mathbb{P}}
\newcommand{\dnormObserved}{\texttt{normal.obs}\xspace}
\newcommand{\dnormSimulated}{\texttt{simNorm.txt}\xspace}
\newcommand{\dunifSimulated}{\texttt{simUnif.txt}\xspace}

\newcommand{\simpleobs}{\texttt{simple.obs}\xspace}

\begin{document}
\title{A Guide to General-Purpose Approximate Bayesian Computation Software}
\author{Athanasios Kousathanas, Pablo Duchen \& Daniel Wegmann\\Department of Biology, University of Fribourg, Switzerland}

\maketitle

\section{Introduction}
There are currently many programs available to conduct ABC analyses (see Table \ref{software}). However, most programs are specific to a particular problem, and the large majority for questions that typically arise in a population-genetics setting.  These include programs to infer demographic histories (e.g. \textit{ONeSAMP} , \textit{PopABC}, \textit{msABC} or \textit{DIYABC}) , to infer F-statistics (\textit{ABC4F}) or to infer parental contributions in an admixture event (\textit{2BAD}). However, there exist also programs for phylogeographic inference (\textit{msBayes}) , Systems Biology (\textit{ABC-SysBio}) or the inference under stochastic differential equations (the MATLAB package \textit{abc-sde}).

A major benefit of such specific programs is that a model parameterization suitable for simulation-based inference has already been worked out and a set of summary statistics informative for the problem identified. However, a particularly powerful aspect of ABC is its application to almost any inference problem and model, and we will thus focus here on general purpose ABC software designed to be helpful in a large array of ABC applications, either to conduct the whole analysis pipeline or at least specific parts of it.

Specifically, we will discuss two similar ABC pipelines, one provided through the command line program \textit{ABCtoolbox} (version 2.0, \cite{wegmann_abctoolbox:_2010}), and the other by means of combining the two R packages \textit{abc} (version 2.0, \cite{Csillery2012}) and \textit{EasyABC} (version 1.4, \cite{jabot_easyabc:_2013}). Both pipelines offer very similar features and build around a similar logic: 1) they both provide utilities to generate a large set of simulations using external simulation software that employs a variety of sampling algorithms, 2) they both offer algorithms to infer parameters from such a set of simulations, 3) they provide tools to conduct model choice from such sets of simulations performed under different models, and finally, 4) they both offer a series of functions to validate estimations as well as model choice. However, these pipelines differ in specific implementations of some algorithms, which we will outline below, as well as the general way users interact with them. 
The 
packages \textit{EasyABC} and \textit{abc} are used within the statistical software environment R and are hence well suited for people writing their
simulation programs in R or for those familiar with the handling of data sets in this environment. In contrast, \textit{ABCtoolbox} is written purely in \texttt{C++} and run from the command line, resulting in generally faster execution and making it particularly well suited when command- line programs are used to conduct simulations or calculate summary statistics.

In the remainder of this chapter, we will walk the reader through the general usage of these two ABC pipelines. In order to give the reader a chance to replicate our analysis easily, we will use rather simple models and always give a detailed description of all settings used, along with all the code required to replicate the described analyses. We will begin with parameter inference, as we believe this is the step for which the programs discussed here will be used most often. However, we will also discuss how to use these pipelines to perform model choice, and how to conduct simulations using existing software.

\begin{table}[!htbp]
\centering
\caption{\small{General and specific purpose ABC software.}}
\begin{tabular}{p{2.5cm} p{6cm} p{5cm}}
\toprule
Software & Purpose & Reference\\
\midrule
\textit{ABCtoolbox} & General & \cite{wegmann_abctoolbox:_2010}\\
\textit{abc} & General & \cite{Csillery2012}\\
\textit{ABC-EP} & General & \cite{Barthelme2011}\\
\textit{ABC\textunderscore distrib} & General & \cite{Beaumont2002}\\
\textit{ABCreg} & General & \cite{Thornton2009}\\
\textit{EasyABC} & General & \cite{jabot_easyabc:_2013}\\
\textit{DIYABC} & Population genetics & \cite{Cornuet2008}\\
\textit{msABC} & Population genetics & \cite{Pavlidis2010b}\\
\textit{ONeSAMP} & Population genetics & \cite{Tallmon2008}\\
\textit{PopABC} & Population genetics & \cite{Lopes2009}\\
\textit{2BAD} & Population genetics & \cite{Bray2010}\\
\textit{ABC4F} & Population genetics & \cite{Foll2008}\\
\textit{msBayes} & Phylogeography & \cite{hickerson_msbayes:_2007}\\
\textit{ABC-SysBio} & Systems Biology & \cite{Liepe2010}\\
\textit{abc-sde} & Stochastic differential equations & \cite{picchini_inference_2014}\\
\bottomrule
\end{tabular}
\label{software}
\end{table}

\section{Toy models}\label{ToyModels}
We will introduce the usage of the program \textit{ABCtoolbox} and the R packages \textit{EasyAbc} and \textit{abc} through their application to the problem of inferring the mean ($\mu$) and variance ($\sigma^2$) of a normal (Model A) and a uniform (Model B) distribution from a random sample. We will then further show how to use these ABC pipelines to distinguish between these models using model choice.

Using such simple models has two major benefits. First, it will allow us to compare the ABC estimates with those obtained from full likelihood solutions. Second, it is straightforward and quick to generate data under these models using just a few lines of code, for instance when using the free statistical programming language R, which we will do here. However, note that we will also discuss how to generate simulations with existing programs in Section \ref{autosim}.

\subsection{Observed data and summary statistics}
We will begin by generating a data set of 100 random samples under Model A (the normal distribution) for which we would like to infer parameters. This is readily done in R as follows:

\begin{lstlisting}[language=R]
sampleSize <- 100;
data.obs <- rnorm(sampleSize, mean=0, sd=1);
\end{lstlisting}

Note that for this test example we do know the true parameters $\theta=(\mu, \sigma^2)$ that we want to estimate. This will allow us to test the accuracy of the estimation.

Next, we will define a function to calculate summary statistics both on the observed as well as simulated data sets. As summary statistics, we will use here the sample mean, variance and median, along with the smallest (min), largest (max) value, the range (max - min) and the first and third quartile (Q1 and Q3). A vector containing these summary statistics is readily generated in R using the following function:

\begin{lstlisting}[language=R]
calc.stats <- function (x){
  S <- c(mean(x), var(x), median(x), range(x), max(x)-min(x),
  quantile(x, probs=c(0.25, 0.75)));
  names(S) <- c("mean", "var", "median", "min", "max", "range", "Q1", "Q3")
  return(S);
}
\end{lstlisting}

This will now allow us to calculate these summary statistics on the observed data set (object data.obs), save them in the variable $S_{obs}$ (object S.obs) and to also save them in a file called \dnormObserved.

\begin{lstlisting}[language=R]
S.obs <- calc.stats(data.obs);
write.table(t(S.obs), file="normal.obs", quote=F, row.names=F);
\end{lstlisting}

While each run will produce different values, we provide the values we obtained and will use in the following in Table \ref{Table_Obstats} to allow for the replication of all our results.

\begin{table}[!htbp]
\centering
\footnotesize
\captionof{table}{Observed statistics of test data set} \label{Table_Obstats}
\begin{tabular}{cccccccc}
\toprule
mean & var & median & min & max & range & Q1 & Q3\\
0.102&1.14&0.0788&-2.02&3.16&5.18&-0.598&0.799\\
\bottomrule
\end{tabular}
\end{table}

\subsection{Generating simulations}
We will next generate a large number (10,000) of simulations with parameter values drawn from prior distributions and calculate the associated summary statistics for each simulation. In order to allow for a direct comparison between the models, we will assume uniform prior distributions for the mean $\mu \sim U[-1, 1]$ and variance $\sigma^2 \sim U[0.1, 4]$ for both models. We also set the internal random number seed generator equal to one so that the reader can reproduce exactly our results. Simulations for the normal model (Model A) are then generated as follows:

\begin{lstlisting}[language=R]
set.seed(1)
nsim <- 10000;
P.normal <- data.frame(mu=runif(nsim, min=-1, max=1), sigma2=runif(nsim, min=0.1 , max=4));
S.normal <- data.frame(matrix(data=0, ncol=length(S.obs), nrow=nsim));
names(S.normal) <- names(S.obs);
for ( i in 1:nsim ) {
  y <- rnorm(sampleSize, mean = P.normal$mu[i], sd=sqrt(P.normal$sigma2[i]));
  S.normal[i,] <- calc.stats(y);
}
write.table(cbind(P.normal, S.normal), file="simNorm.txt", quote=F, row.names=F);
\end{lstlisting}

Again, we saved the simulations also in a text file (\dnormSimulated) in order to use them with \textit{ABCtoolbox}. Note that \textit{ABCtoolbox} requires the parameters and statistics in the same file, which is achieved by binding the data frames containing the parameters (\texttt{P.normal}) and statistics (\texttt{S.normal}) together using \texttt{cbind()}.

Generating simulations under the uniform model (Model B) is achieved similarly. However, since the R function \texttt{runif()} requires the two limits $a,b$ of the uniform distribution, rather than the mean and variance, we need to calculate them from the parameters $\mu$ and $\sigma^2$ after each draw as $a=\mu - \sqrt{3\sigma^2}$ and $b=\mu + \sqrt{3\sigma^2}$, respectively.

\begin{lstlisting}[language=R]
nsim <- 10000;
P.unif <- data.frame(mu=runif(nsim, min=-1, max=1), sigma2=runif(nsim, min=0 , max=4));
S.unif <- data.frame(matrix(data=0, ncol=length(S.obs), nrow=nsim));
names(S.unif)<-names(S.obs);
for ( i in 1: nsim ) {
  y <- runif(sampleSize, min=P.unif$mu[i]-sqrt(3*P.unif$sigma2[i]), max=P.unif$mu[i]+sqrt(3*P.unif$sigma2[i]));
  S.unif[i,] <- calc.stats(y);
}
write.table(cbind(P.unif, S.unif), file="simUnif.txt", quote=F, row.names=F);
\end{lstlisting}

\section{Parameter inference} \label{inference}
The estimation of posterior distributions is straightforward with both \textit{ABCtoolbox} and the R package \textit{abc}. As a common feature, they both implement the basic rejection algorithm originally introduced by \cite{Tavare1997} and \cite{Pritchard1999}, but they differ in the post-sampling adjustment algorithms they offer. Specifically, the package \textit{abc} implements the original post-sampling adjustment based on a local linear regression initially introduced by \cite{Beaumont2002}, as well as an extension to non-linear models with heteroscedastic variance \citep{Blum2010}. In contrast, \textit{ABCtoolbox} offers an implementation of the general linear model adjustment algorithm (ABC-GLM)\index{General linear model adjustment (ABC-GLM)} introduced by \cite{leuenberger_bayesian_2010}. In this section, we will discuss differences between these algorithms as well as how to use them in the present example.

\subsection{Rejection Algorithm}

\paragraph{\textit{abc}}\index{abc (R package)}
To start using \textit{abc}, the package has first to be installed and loaded in R, which is done by the following two commands:

\begin{lstlisting}[language=R]
install.packages("abc");
library(abc);
\end{lstlisting}

To now conduct an ABC rejection on the data simulated under the normal model (Model A), simply use the function \texttt{abc()} with the argument \texttt{method="rejection"} and by specifying the tolerance to be applied.

\begin{lstlisting}[language=R]
rejection <- abc(S.obs,P.normal,S.normal,tol=0.01,method="rejection");
\end{lstlisting}

Here, \texttt{S.obs}, \texttt{P.normal} and \texttt{S.normal} refer to the vector of observed summary statistics $S_{obs}$ and the data frames containing the simulated parameters and summary statistics, respectively, as generated under Section \ref{ToyModels}. The additional argument \texttt{tol} specifies the fraction of simulations to be retained based on their distance to the observed summary statistics. A \index{Tolerance}tolerance of $0.01$, for example, indicates that the posterior density will be estimated from the parameter values of the 1\% of all simulations that produced summary statistics closest to the observed summary statistics based on an euclidean distance metric.

The package \textit{abc} offers an internal plotting function \texttt{hist.abc()} to display posterior distributions. Since this function overloads the basic \texttt{hist()} function of R, it can be called on an \texttt{abc} object by simply typing:

\begin{lstlisting}[language=R]
hist(rejection);
\end{lstlisting}

Alternatively, it is also possible to use general R functions such as \texttt{hist()} or \texttt{density()} to plot posterior distributions. The results we obtained for the observed data shown above is plotted using \texttt{density} in Figure \ref{Figure_posteriors}.

\paragraph{\textit{ABCtoolbox}}\label{parestABCtoolbox}\index{ABCtoolbox}
In contrast to the R packages discussed here, the program \textit{ABCtoolbox} is a program to be used from the command line, preferentially in a Unix/Linux environment. While it can be run from the Windows command prompt, it is recommend to use the \textit{cygwin} Unix-like interface on a Windows computer to benefit from all features of \textit{ABCtoolbox}.

\textit{ABCtoolbox} accepts input settings both directly from the command line, as well as through an input file. A list of all arguments relevant for estimation and discussed in this chapter is provided  in Table \ref{est_settings}. To perform parameter estimation for the normal distribution example, for instance, the following input file may be used:

\begin{lstlisting}
task estimate
simName '*\dnormSimulated*'
obsName '*\dnormObserved*'
params 1-2
maxReadSims 10000
numRetained 100
writeRetained
maxCor 1.0
\end{lstlisting}

Here, the argument \texttt{task} is set to \texttt{estimate} in order to run \textit{ABCtoolbox} in estimation mode. Then, the argument \texttt{simName} specifies the name of the file containing the performed simulations. This file is requested to contain the used model parameter values together with the associated statistics, the names of which are provided in the first line. Similarly, the file specified with the argument \texttt{obsName} should contain the summary statistics $S_{obs}$ calculated from the observed data, again with the first line of the file containing the names of the statistics and the second line the associated values. Using the argument \texttt{params}, \textit{ABCtoolbox} is further told which columns of the simulation file contain the model parameters to be estimated. Note that the simulation file may contain an arbitrary number of additional columns that will be ignored if they are neither specified to be model parameters with the argument \texttt{params} nor summary statistics also
present in the file with the observed summary statistics.

The additional required arguments \texttt{maxReadSims} and \texttt{numRetained} specify the maximum number of lines (simulations) that will be read from the simulation file, and the number of simulations to be retained in the rejection step, respectively. Finally, the argument \texttt{maxCor} specifies the maximal allowed correlation between summary statistics. If we also add the argument \texttt{pruneCorrelatedStats} then the analysis will be performed by using statistics that their pairwise correlation do not exceed the \texttt{maxCor} threshold \index{Correlated statistics}. We will discuss the issues with correlated statistics below (section \ref{statsSummary}), but set this option here to 1 in order to include all statistics in the calculations and to avoid \textit{ABCtoolbox} complaining about the presence of highly correlated statistics in our toy models.

To run \textit{ABCtoolbox} with this input file (assuming it was saved under the name \texttt{estimate.input}), simply run in the command:

\begin{lstlisting}[language=bash]
./ABCtoolbox estimate.input
\end{lstlisting}

Alternatively, the example can be run without using an input file by specifying the commands in the command line as:

\begin{lstlisting}[language=bash]
./ABCtoolbox task=estimate simName='*\dnormSimulated*' obsName='*\dnormObserved*'
params=1-2 maxReadSims=10000 numRetained=100 writeRetained maxCor=1.0
\end{lstlisting}

The output of such a run is found in a series of files, the names of which begin with a prefix that can be set with the argument \texttt{outputPrefix}, a tag referring to its content, and a number referring to the data set and model for which the estimation has been conducted. While an exhaustive list of all output filename tags discussed in this chapter is given in Table \ref{Table_posteriorfiles}, we will focus here on the file with tag \texttt{BestSimsParamStats}, which contains the retained simulations and is used to plot the rejection posterior.

The posterior estimates for \textit{$\mu$} and \textit{$\sigma^2$} obtained from the ABC-rejection algorithm are readily plotted in R using the \texttt{density()} function.

\begin{lstlisting}[language=R]
par(mfrow=c(1,2))
ABCrej<-read.delim("ABC_GLM_model0_BestSimsParamStats_Obs0.txt",sep="\t");
plot(density(ABCrej$mu,from=-1,to=1),main=expression(mu));
plot(density(ABCrej$sigma2,from=0,to=4),main=expression(sigma^2));
\end{lstlisting}

The results we obtained for the observed data shown above are plotted using \texttt{density} in Figure \ref{Figure_posteriors}. 

\subsection{Post-sampling adjustments}\index{Post-sampling adjustments}
Posterior distributions estimated with the rejection algorithm tend to be much broader than the true posterior distributions. This is shown for the normal model in Figure \ref{Figure_posteriors}, but has been observed generally and is due to the often relatively large distance thresholds leading to parameter values resulting in summary statistics rather distant from $S_{obs}$ to be accepted. Obviously, this loss of precision can be reduced by being more restrictive in accepting simulations, but this may require unrealistically computational efforts, particularly in more complex models.

An alternative is to correct for the effect of using large thresholds by exploiting the often simpler relationship between model parameters and summary statistics locally around the observed summary statistics. In a landmark paper, \citet{Beaumont2002} assume a linear relationship between model parameters and summary statistics locally among the retained simulations and proposed to use this relationship to project the parameter values of all retained parameter values to $S_{obs}$. More recently,  \citet{Blum2010} introduced an extension of this approach by fitting a non-linear, heteroscedastic model using neural networks. Both of these algorithms are implemented in the R package \textit{abc}.

In contrast, \textit{ABCtoolbox} offers an implementation of the ABC-GLM algorithm\index{General linear model adjustment (ABC-GLM)} introduced by \citet{leuenberger_bayesian_2010} that estimates a local likelihood function instead of directly targeting the posterior distribution. While potentially slightly slower, this formulation is flexible in the choice of prior distributions and allows for model choice based on the marginal density. In practice, however, all mentioned post-sampling adjustment algorithms tend to give very similar results and the reader is advised to validate any estimation carefully in which these algorithms produce diverging estimates.

\paragraph{\textit{abc}}
The two post-sampling adjustments implemented in the R package \textit{abc} are used by simply choosing the appropriate \texttt{method} when calling the \texttt{abc()} function. There are three different methods available: \texttt{loclinear}, \texttt{ridge} and \texttt{neuralnet}, which correspond, respectively, to the classic regression adjustment introduced by \citet{Beaumont2002}\index{Regression adjustment}\index{Post-sampling adjustments}, a version of this algorithm using a ridge regression to deal with extensive collinearity among statistics\index{Ridge regression}\index{Post-sampling adjustments}, and the non-linear, heteroscedastic regression proposed by \citet{Blum2010}\index{Neural network adjustment}\index{Post-sampling adjustments}. When using the loclinear method\index{Loclinear method}, if a warning appears regarding the collinearity of the design matrix then we recommend to use the ridge method instead.

The following commands will perform posterior estimation using these algorithms on the toy model introduced above.

\begin{lstlisting}[language=R]
regression <- abc(S.obs,P.normal,S.normal,tol=0.01,method="loclinear");
neural <- abc(S.obs,P.normal,S.normal,tol=0.01,method="neuralnet");
\end{lstlisting}

The built-in function \texttt{hist()} can then again be used to plot the estimated posterior distributions.

\begin{lstlisting}[language=R]
hist(regression);
hist(neural);
\end{lstlisting}

Another function provided by the package \textit{abc} is \texttt{plot.abc}, which can be used to plot the densities of the estimated posterior distributions together with additional, informative plots such as the prior distribution, the distribution of euclidean distances, and the residuals of the regression. Since this function overloads the standard R function \texttt{plot()}, it is simply used as follows:

\begin{lstlisting}[language=R]
plot(regression,param=P.normal);
plot(neural,param=P.normal);
\end{lstlisting}

Alternatively, the estimated posterior distributions can also be plotted using the R function \texttt{density}. For that purpose, one has to access specific elements of the object returned by the \texttt{abc()} function, namely the projected model parameter values as \texttt{adj.values} as well as their weights \texttt{weights}. The following R commands, for instance, plot the posterior densities obtained via the regression and neural network adjustment\index{Neural network adjustment}\index{Post-sampling adjustments} for $\mu$:

\begin{lstlisting}[language=R]
plot(density(regression$adj.values[,1], weights=regression$weights/sum(regression$weights)),main=expression(mu));
plot(density(neural$adj.values[,2], weights=neural$weights/sum(neural$weights)),main=expression(sigma^2));
\end{lstlisting}

Posterior distributions plotted using these functions are compared to those obtained through other methods in Figure \ref{Figure_posteriors}. Note that the object returned also contains the retained model parameter values in the element \texttt{unadj.values} that can be used to plot the rejection posterior distribution.

\begin{lstlisting}[language=R]
plot(density(regression$unadj.values[,1]),main=expression(mu));
plot(density(regression$unadj.values[,2]),main=expression(sigma^2));
\end{lstlisting}

\paragraph{\textit{ABCtoolbox}}\label{ABCGLM}\index{ABCtoolbox}
When running \textit{ABCtoolbox} in \texttt{estimation} mode, the \index{General linear model adjustment (ABC-GLM)}\index{Post-sampling adjustments} ABC-GLM adjustment  introduced by \cite{leuenberger_bayesian_2010} is performed automatically and the results available in the output file with tag \texttt{MarginalPosteriorDensities}. To plot the posterior estimates in R, simply load that file and use the function \texttt{density}.

\begin{lstlisting}[language=R]
par(mfrow=c(1,2))
ABCglm <- read.delim("ABC_GLM_model0_MarginalPosteriorDensities_Obs0.txt");
plot(ABCglm$mu,ABCglm$mu.density,type="l");
plot(ABCglm$sigma2,ABCglm$sigma2.density,type="l");
\end{lstlisting}

\begin{table}[!htbp]
\footnotesize
\centering
\captionof{table}{\textit{ABCtoolbox} settings for estimation.} \label{est_settings}
\begin{tabulary}{1.0\textwidth}{p{2cm}p{4cm}p{8cm}}
\toprule
Setting type&Setting&Description\\
\cmidrule{1-3}
\multirow{16}{*}{Basic}&task&Task to be performed. Possible options are: simulate, estimate, findStatsModelChoice\\
&estimationType&Standard estimation performs the GLM approach \cite{leuenberger_bayesian_2010}. \\
&params&Specify the parameter columns in the file that contains the simulations.\\
&simName&File containing simulations.\\
&obsName& File containing observed summary statistics.\\
&numRetained& No. of simulations to retain.\\
&maxReadSims& Maximum number of read simulations.\\
&pruneCorrelatedStats& Remove statistics that are correlated, possible options 0 (retain) or 1 (remove).  \\
&maxCor&Maximum acceptaple correlation coefficient between statistics. \\
&outputPrefix&Prefix for output files.\\
&writeRetained&Indicate whether to write retained simulations which can be used to obtain ABC rejection posteriors.\\
&standardizeStats&Standardize statistics.\\\cmidrule{1-3}
\multirow{5}{*}{Validation}&obsPValue&The number of retained data sets for testing how well the inferred GLM model fits the observed data in multidimensional space (section \ref{modValwrong}).\\
&tukeyPValue& The number of retained data sets for performing the Tukey test (section \ref{modValwrong}).\\
&modelChoiceValidation&The number of cross-validation replicates for validating model choice (section \ref{modVal}).\\
&randomValidation&The number of cross-validation replicates for random parameter validation (section \ref{parValbias}).\\
&retainedValidation&The number of cross-validation replicates for retained parameter validation (section \ref{parValbias}).\\\cmidrule{1-3}
\multirow{4}{*}{Posterior density}&posteriorDensityPoints&Number of points to estimate posterior density.\\
&diracPeakWidth&Smoothing parameter for posterior densities.\\
&jointPosteriors&Comma separated list of parameters for which the joint posterior is to be evaluated (section \ref{multidimpost}).\\
&jointPosteriorDensityPoints&No. of points to evaluate joint posterior (section \ref{multidimpost}).\\

\bottomrule
\end{tabulary}
\end{table}

\begin{table}[!htbp]
\centering
\captionof{table}{\textit{ABCtoolbox} estimation output files.} \label{Table_posteriorfiles}
\footnotesize
\begin{tabulary}{1.0\textwidth}{p{3.5cm}p{5cm}p{7cm}}
\toprule
File type&File tag&Content\\\cmidrule{1-3}
\multirow{5}{*}{Basic}&MarginalPosteriorCharacteristics&Characteristics of marginal posterior distributions (e.g., mode, mean, quantiles).\\
&BestSimsParamStats&Retained simulations from ABC-rejection.\\
&MarginalPosteriorDensities&GLM-adjusted marginal posterior densities.\\
&jointPosterior&GLM-adjusted joint posterior estimates.\\
&modelFit&Model choice results including bayes factors and posterior support for compared models.\\\cmidrule{1-3}
\multirow{2}{*}{Parameter validation}&RandomValidation&Results from random validation.\\
&RetainedValidation&Results from retained validation.\\\cmidrule{1-3}
\multirow{1}{*}{Model choice validation}&modelChoiceValidation&Results for model choice validation.\\
\bottomrule
\end{tabulary}
\end{table}

\subsection{Multidimensional posteriors}\label{multidimpost}\index{Multidimensional posteriors}

\begin{figure}
\centering
\includegraphics[keepaspectratio,scale=0.5]{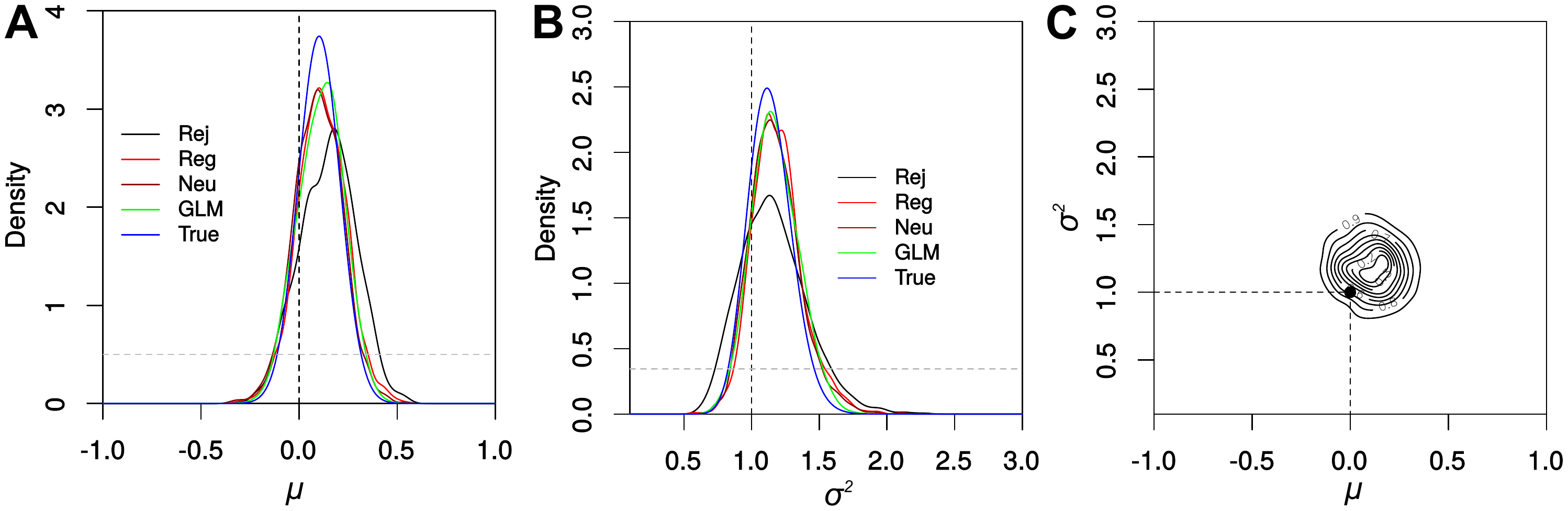}
\caption{\label{Figure_posteriors} Posterior densities for the mean (A) and variance (B) parameters of the normal distribution model produced by a variety of methods. Dashed grey lines indicate prior densities. (C) Joint posterior density for \textit{$\mu$} and \textit{$\sigma^2$} produced with \textit{ABCtoolbox}. Dashed black lines indicate true parameter values for all panels.
}

\end{figure}

\paragraph{\textit{abc}}\index{abc (R package)}
Apart from marginal posterior distributions, both the R package \textit{abc} as well as \textit{ABCtoolbox} are capable of estimating multidimensional posterior distributions. In the case of \textit{abc}, however, the posterior densities have to be estimated using standard functions of R such as \texttt{kde2d()} for two-dimensional posterior distributions. In our toy model, for instance, using

\begin{lstlisting}[language=R]
posterior2d <- kde2d(regression$adj.values[,1],regression$adj.values[,2],n=100);
contour(posterior2d,xlab=expression(mu),ylab=expression(sigma^2));
\end{lstlisting}

where, \texttt{regression\$adj.values[,1]} represents the projected model parameter values for $\mu$, \texttt{regression\$adj.values[,2]} represents the projected model parameter values for $\sigma^2$, and \texttt{n=100} specifies the number of marginal grid points to be used for the density estimation. These R commands will produce a plot similar to the one in figure \ref{Figure_posteriors}C.

\paragraph{\textit{ABCtoolbox}}\index{ABCtoolbox}\index{Post-sampling adjustments}
To generate multidimensional posterior densities on a grid, simply add the argument \texttt{jointPosteriors}, followed by the names of the model parameters for which the multidimensional posterior distribution is to be estimated. In addition, the number of marginal grid points has to be specified using \texttt{jointPosteriorDensityPoints}. The joint posterior estimates for the parameters \textit{$\mu$} and \textit{$\sigma^2$} of the normal distribution model,  for instance, is thus estimated by simply running \textit{ABCtoolbox} with a modified \texttt{estimate.input} input file that contains these two additional lines:

\begin{lstlisting}
jointPosteriors mu,sigma2
jointPosteriorDensityPoints 100
\end{lstlisting}

Note that the total number of grid points grows exponentially with the dimensionality. In this above example, the density will be evaluated at $100 \times 100 = 10^4$ positions. Running such a command for a four dimensional posterior will already result in $10^8$ positions at which the density has to be estimated.

When running \textit{ABCtoolbox} with the additional arguments \texttt{jointPosteriors} and \texttt{jointPosterior DensityPoints}, the posterior density at each grid point will be written to an output file with tag \texttt{jointPosterior}. The resulting joint posterior of \textit{$\mu$} and \textit{$\sigma^2$} can then be plotted in R using the function \texttt{contour()}.

\begin{lstlisting}[language=R]
plot2D <- read.delim("ABC_GLM_model0_jointPosterior_1_2_Obs0.txt");
x <- unique(plot2D$mu);S.unif$var
y <- unique(plot2D$sigma2);
z_density <- matrix(data=plot2D$density,nrow=length(x),ncol=length(y),byrow=F);
contour(x,y,z_density,xlab=expression(mu),ylab=expression(sigma^2));
\end{lstlisting}

Since densities may be hard to interpret, \textit{ABCtoolbox} also calculates and prints the smallest high posterior density interval (HDI) \index{High posterior density interval (HDI)}including each grid point to the same output file. The HDI corresponds to a posterior credible interval \index{Credible interval} in the multidimensional parameter space and hence allows the generation of contour plots where contour lines indicate posterior credible intervals as follows:

\begin{lstlisting}[language=R]
z_HPD <- matrix(data=plot2D$HDI,nrow=length(x),ncol=length(y),byrow=F);
contour(x,y,z_HPD,xlab=expression(mu),ylab=expression(sigma^2));
\end{lstlisting}

The two dimensional posterior distribution we obtained this way for the normal distribution example is given in Figure \ref{Figure_posteriors}C.

Note that using a grid evaluation is not suitable to estimate multidimensional densities in high dimensions as the total number of grid points grows exponentially with the dimensionality. In this above example, the density will be evaluated at $100 \times 100 = 10^4$ positions. Running such a command for a four dimensional posterior will already result in $10^8$ positions at which the density has to be estimated. An alternative is to generate samples from the joint posterior distribution from which densities are estimated using kernel estimators. To generate samples from high-dimensional posterior distributions after post-sampling adjustment, \textit{ABCtoolbox} also implements an MCMC algorithm. While we will not discuss that algorithm here, we refer the user to the manual of \textit{ABCtoolbox} for more details.

\subsection{Validation of parameter estimation}\label{parVal}\index{Validation}

\subsubsection{Using a wrong model}\label{modValwrong}\index{Validation}
An essential first validation step is to check whether the observed statistics can be reproduced by the examined model. A failure of the model to reproduce some of the statistics may indicate that a model is either not reflecting reality close enough, or that inappropriate prior distributions have been used (e.g., too narrow distributions). More importantly, all post-sampling adjustments assume that the model fitted to the model parameters and summary statistics can be used to either accurately project retained simulations to $S_{obs}$ (the methods implemented in \textit{abc}), or is an accurate description of the likelihood of $S_{obs}$ with the parameter range of the retained simulations (the method implemented in \textit{ABCtoolbox}). A violation of these assumptions leads to an extrapolation to an area of the summary statistics space for which no samples have been obtained, and is hence is prone to biased inference.

\begin{figure}[!htbp]
\centering
\includegraphics[keepaspectratio,scale=0.7]{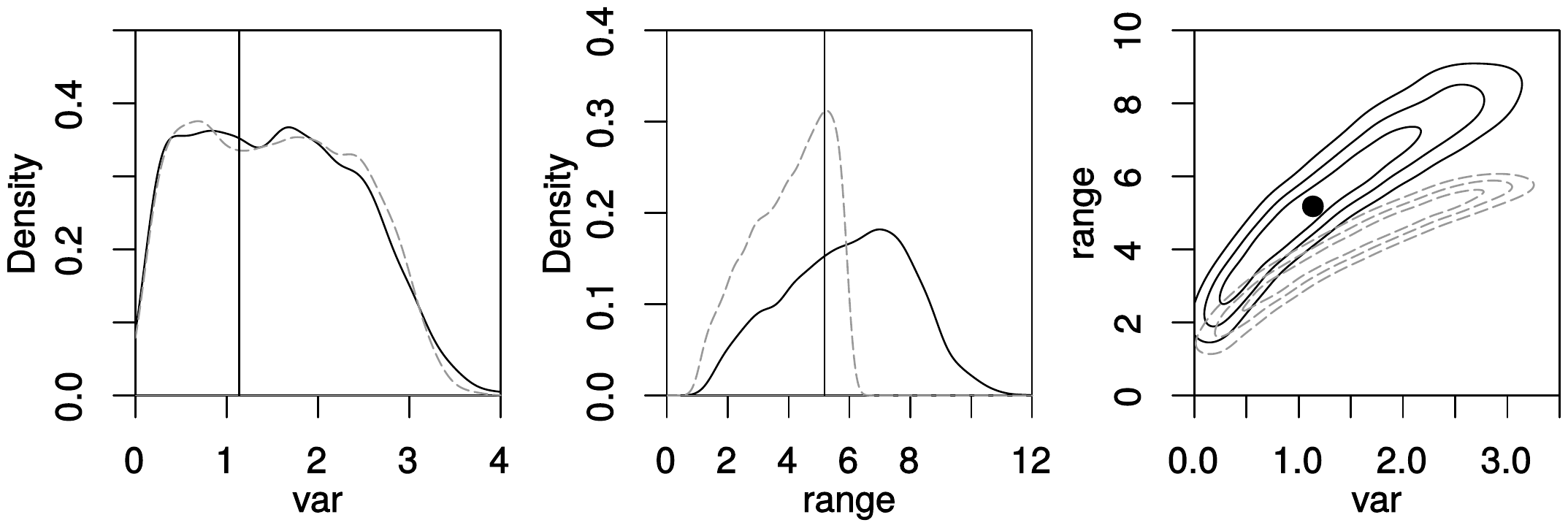}
\caption{\label{Figure_dist} The distributions of simulated versus observed statistics \texttt{variance} (left panel) and \texttt{range} (middle panel) and their joint distribution (right panel) for the normal (black) and uniform (grey) distribution models. Observed data is shown with a black vertical line in the left and middle panels and with a black dot in the right panel.
}
\end{figure}

Checking if the observed summary statistics $S_{obs}$ are within the range of summary statistics generated by the model is, however, a bit tricky in higher dimensions. For instance, consider the marginal summary statistics distributions shown in Figure \ref{Figure_dist} for the normal and uniform toy models, respectively. These distributions were plotted in R using the \texttt{density()} function directly from the simulated data.

\begin{lstlisting}[language=R]
plot(density(S.normal$var), col='black');
lines(density(S.unif$var), col='grey');
abline(v=S.obs[2])
\end{lstlisting}

These plots suggest that both summary statistics are readily generated by both models.  However, a mismatch might manifest when looking at combinations of summary statistics. To check this we can plot two-dimensional distributions of pairs of summary statistics using the R functions \texttt{kde2d} and \texttt{contour}.

\begin{lstlisting}[language=R]
d <- kde2d(S.unif$var, S.unif$range, n=100);
contour(d$x, d$y, d$z);
\end{lstlisting}

As is shown in the rightmost panel in Figure \ref{Figure_dist}, the combination of the two observed summary statistics \texttt{variance} and \texttt{range} can indeed not be reproduced by the uniform model. However, this fact is not visible when looking at marginal densities only.

\paragraph{\textit{ABCtoolbox}}
Since visual inspection is only fruitful for a limited number of dimensions, \textit{ABCtoolbox} offers two statistical tests for assessing whether a given model can reproduce the observed data $S_{obs}$ in the multidimensional statistics space. The first test compares the marginal density (also called marginal likelihood) of the observed data to the marginal density of the retained simulations. The fraction of retained simulations with smaller or equal marginal density than the observed data is then provided as \textit{the marginal density P-value}\index{Marginal density P-value} where small values indicate a poor fit of the model to the observed data.

The second test evaluates how central the observed data lies within the multidimensional cloud of retained simulations by reporting the fraction of retained simulations with smaller or equal \index{Tukey depth}\index{Tukey P-value}Tukey depth than the observed data as the \textit{Tukey P-value} (\cite{cuesta-albertos_random_2008,Adrion2014}). The Tukey depth is a common measure of centrality analogous to the median in one dimension and is calculated by \textit{ABCtoolbox} for a retained simulation (or the observed data) as the smallest fraction of retained simulations which can be separated from the rest of the simulations using a hyper plane through the chosen simulation (or the observed data). Again, a low Tukey P-value indicates a poor fit of the model since the observed data appears to be at the periphery of the retained cloud. However, note that the opposite is not necessarily true. Indeed, even a poor model  (e.g. a model producing summary statistics at random) may be capable of generating a cloud of summary statistics surrounding $S_{obs}$, 
and will thus pass both tests.

To perform these tests, simply call \textit{ABCtoolbox} with the arguments \texttt{marDensPValue} and \texttt{tukeyPValue}, where each of them indicates the number of retained simulations to be used when calculating the respective P-value. When adding the following two lines to the input file \texttt{estimate.input}, for instance, \textit{ABCtoolbox} will use 1,000 retained simulations to evaluate the P-values.

\begin{lstlisting}
marDensPValue 1000
tukeyPValue 1000
\end{lstlisting}

The results we obtained for these tests for our toy models is shown in Table \ref{tab:Pvalues}. As expected from the visual inspection in Figure \ref{Figure_dist}, the uniform model is not capable of reproducing the observed data and hence fails both tests.

\begin{table}[!htbp]
\centering
\captionof{table}{Observed P-value and Tukey P-value results for normal distribution example} \label{tab:Pvalues}
\begin{tabular}{c c c c c}
\toprule
Model & marginal density & marginal density P-Value&Tukey Depth&Tukey P-Value\\\hline
1& 1158.75 & 0.098&0.13&0.96\\
2& $4.16\times10^{-12}$ & 0 & 0 & 0 \\
\bottomrule
\end{tabular}
\end{table}

\subsubsection{Cross-validation / Accuracy of point estimates}\label{parValacc}\index{Cross-validation}\index{Validation}
The accuracy of posterior point estimates is generally assessed by estimating the parameters for data sets for which the true parameter values are known. This is readily done in an ABC setting as a leave-one-out test in which one of the provided simulations is randomly chosen and all other are used to infer the parameter estimates for this data (often called ``pseudo-observed'' data). The inferred posterior point estimates such as the maximum a posteriori (MAP or posterior mode), the posterior mean or the posterior median are then plotted against the parameter values used to generate the data (referred to as the ``true parameters''). This process (also called cross-validation) is then repeated for many ``pseudo-observed'' data sets to obtain a general measure of accuracy. The procedure may also be repeated to test specific ABC settings such as the effect of the choice of tolerance or the number of available simulations.

\paragraph{\textit{abc}}\index{abc (R package)}
To use this cross-validation algorithm with the R package \textit{abc}, simply call the function \texttt{cv4abc()} with the arguments matching those of the estimation plus the additional argument \texttt{nval}, which specifies how many pseudo-observed data sets are to be used. However, note that the observed data does not have to be provided, as it is not used in cross validation. The following code, for instance, will conduct cross validation on the normal distribution example for the neural network estimation algorithm based on 100 individual pseudo-observed data sets.

\begin{lstlisting}[language=R]
cv.neural <- cv4abc(P.normal, S.normal, tols=0.1, statistic="mode",method="neuralnet", nval=100);
\end{lstlisting}

Such a call will return an object containing both the true parameter values (element \texttt{true}) as well as the estimated parameter values (element \texttt{estim}), which can be used to estimate correlations among them and plot for visual inspection. A plot such as the one shown in Figure \ref{Figure_parValid1} is generated by

\begin{lstlisting}[language=R]
plot(cv.neural);
\end{lstlisting}

\paragraph{\textit{ABCtoolbox}}\label{parValABCtoolbox}\index{ABCtoolbox}
\textit{ABCtoolbox} offers two flavors of this cross-validation algorithm: either by picking simulations randomly, or by picking simulations only among those that were retained. The former, which is invoked through the argument \texttt{randomValidation},  corresponds to picking parameter values from the prior distribution and is thus informative above the overall accuracy of the ABC estimation under the chosen model. The latter, which is invoked using the argument \texttt{retainedValidation}, is informative about the accuracy of the estimation for the parameter space leading to similar data as the one observed.

\begin{lstlisting}
task estimate
simName '*\dnormSimulated*'
obsName '*\dnormObserved*'
params 1-2
maxReadSims 10000
numRetained 1000
maxCor 1.0
randomValidation 1000
retainedValidation 1000
\end{lstlisting}

When running \textit{ABCtoolbox} with one or both of those arguments, an additional output file with tag \texttt{randomValidation} or \texttt{retainedValidation} is generated that contains the true parameter values along with a series of posterior point estimates for each parameter, namely the MAP (or mode) as well as the posterior mean and median. This file can then be loaded into R to generate plots such as those shown in Figure \ref{Figure_parValid1} as follows:

\begin{figure}[!htbp]
\centering
\includegraphics[keepaspectratio,scale=0.6]{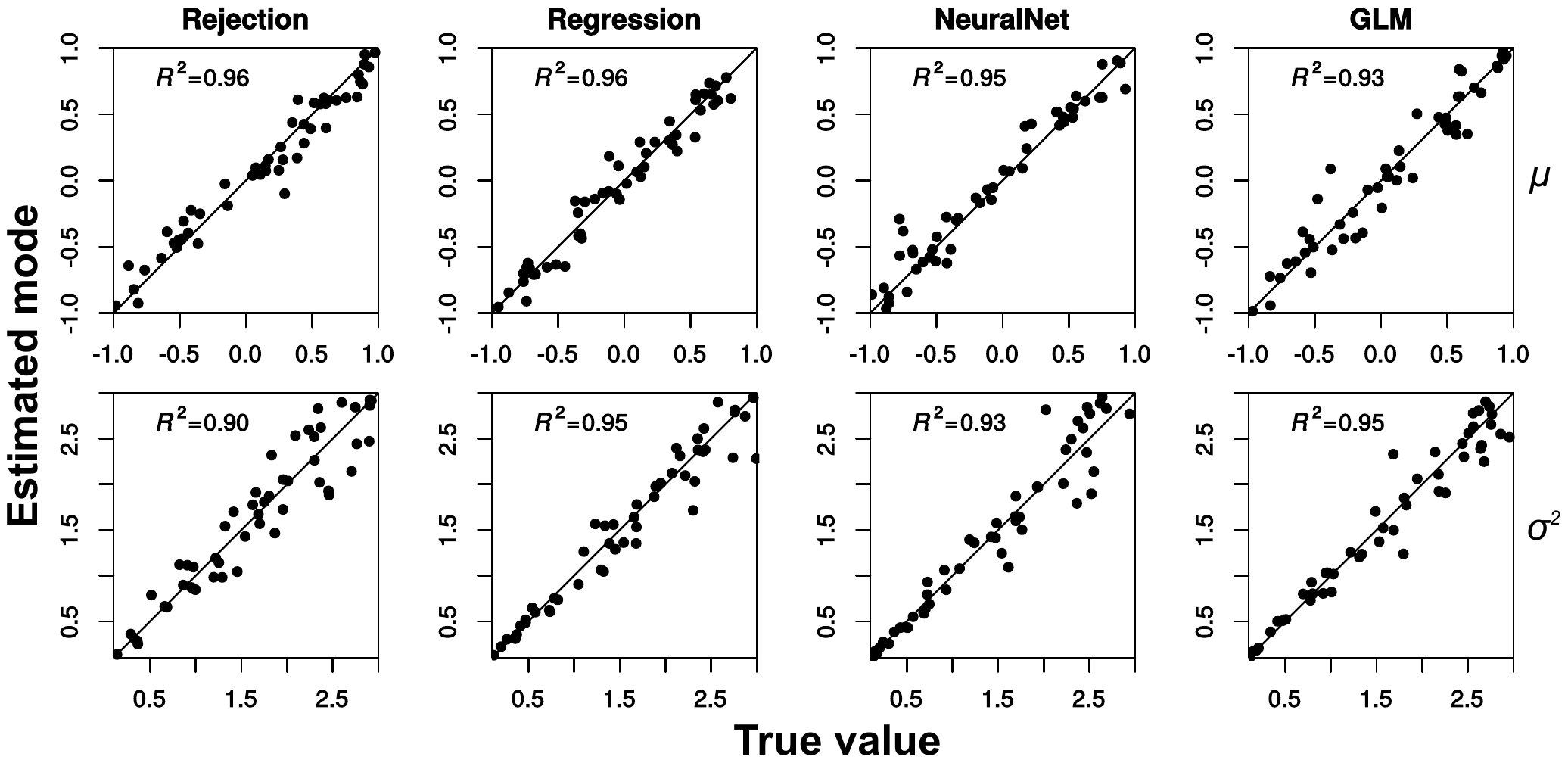}
\caption{\label{Figure_parValid1} Parameter validation using different methods. The estimated posterior mode for mean and variance is plotted against the true values.}
\end{figure}

\begin{lstlisting}[language=R]
Random_validation <- read.delim("ABC_GLM_model0_RandomValidation.txt");
Retained_validation <- read.delim("ABC_GLM_model0_RetainedValidation_Obs0.txt");
plot(Random_validation$mu, Random_validation$mu_mode);
plot(Retained_validation$mu, Retained_validation$mu_mode);
\end{lstlisting}

\subsubsection{Checking for biased posteriors}\label{parValbias}\index{Posterior bias}\index{Validation}
Pseudo-observed data sets can be used equally to detect potential biases in the marginal posterior distributions. If the posterior distributions of a parameter were unbiased, the position of the true parameters across many replicates must be given by the posterior densities. We proposed to test this directly using the \index{Probability Integral Transform Test (PIT)}probability integral transform test (PIT histogram, or coverage property) \cite{wegmann_efficient_2009,Prangle2014}. This is done by recording the position of the true parameter value in the cumulative posterior distribution (the posterior quantile) for each pseudo-observed data set. In case posteriors were unbiased, these posterior quantiles must be uniformly distributed between 0 and 1. Similarly, the smallest \index{High posterior density interval (HDI)}high posterior density intervals (HDI) containing the true parameter value must be distributed uniformly.

\begin{figure}[!htbp]
\centering
\includegraphics[keepaspectratio,scale=0.4]{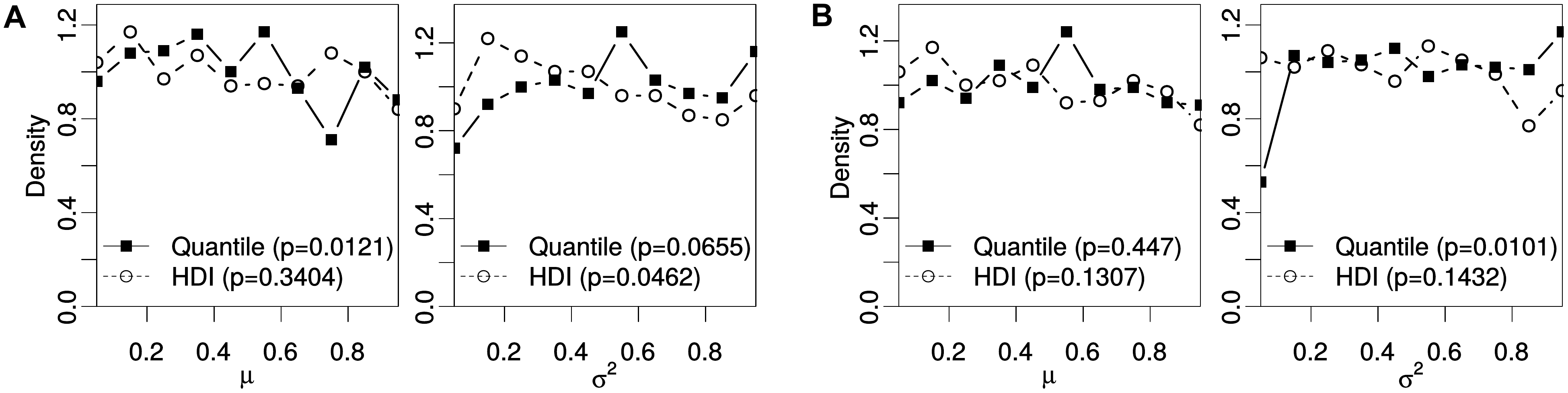}
\caption{\label{Figure_parValid2} Parameter validation testing uniformity of the distribution of posterior quantiles and HDI when using random (A) and retained (B) simulations.}
\end{figure}

\paragraph{\textit{ABCtoolbox}}
The procedure for this is the same as the one described in section \ref{parValABCtoolbox}. The same output file will contain information that allows us to check for biased posteriors. For instance, the output from these analyses can be used to determine whether the posterior quantiles and HDI are uniformly distributed either by visual inspection or by performing a statistical test such as the Kolmogorov-Smirnov test: 

\begin{lstlisting}[language=R]
hist(Random_validation$mu_HDI)
ks.test(Random_validation$mu_HDI,"punif")
\end{lstlisting}

The result of the random and retained validation analyses for the normal distribution example can be seen in Figure \ref{Figure_parValid2}A and Figure \ref{Figure_parValid2}B, respectively.

\section{Model choice}\index{Model choice}
While model choice is commonly used in Bayesian statistics, it is contentious in an ABC setting due to the problem that even summary statistics sufficient for both models may lead to biased inferences \citep{Didelot2011,robert_lack_2011}. Nonetheless, ABC model choice has been used successfully in practice and both ABC pipelines discussed here offer algorithms to conduct model choice when simulations from multiple models are available. However, the user is advised to validate any ABC model choice carefully and we will discuss here tools provided by the R package \textit{abc} as well as \textit{ABCtoolbox} to aid in that crucial step.

\subsection{Inferring Bayes Factors}\index{Bayes factors}
Bayesian model choice relies on the estimation of model posterior probabilities or Bayes factors (the ratios of posterior probabilities of competeing models). In standard Bayesian statistics, these are estimated from the marginal densities (or marginal likelihood) of the compared models, where the marginal density of model $i$ is defined as the integral of the likelihood function, weighted by the prior distribution:

\begin{equation*}
\P(M_i|D) = \int \P(D|\theta_i, M_i )\P(\theta_i|M_i) d\theta_i.
\end{equation*}

Importantly, the marginal density is thus affected by the choice of prior distribution, and hence what is evaluated in a Bayesian setting is thus the combination of a stochastic model and the prior destribution specified on its parameters.

In case the likelihood function is not available for analytical evaluation, both posterior probabilities and Bayes factors can be estimated using ABC. The algorithm implemented in \textit{ABCtoolbox}, for instance, fits a likelihood model to the retained data, and then uses this model to analytically calculate the marginal density for each model. In contrast, the algorithms available in the R package \textit{abc} use the fact that  the marginal density is proportional to the fraction of simulations that resulted in simulations close to the observed data $S_{obs}$ when the simulations are generated according to the prior distribution. The posterior probabilities of the different models are thus estimated from the relative number of simulations being close to $S_{obs}$ either from direct counting or through a regression adjustment similar to parameter inference. Note that it is crucial for both algorithms that the exact same summary statistics have been calculated under both models.

To illustrate the model choice algorithms implemented, we will attempt to estimate which of the two toy models introduced above (the normal and uniform model) was used to generate the observed data. We refer the reader to section \ref{ToyModels} for more details on those models.

\paragraph{\textit{abc}}\index{abc (R package)}
In order to conduct model choice\index{Model choice} with \textit{abc}, the summary statistics of all models have to be concatenated into a single data frame or matrix. In addition, a vector indicating for each simulation the model under which it has been generated has to be created. For our toy models, this is simply achieved as follows:

\begin{lstlisting}[language=R]
allSimulations <- rbind(S.normal,S.unif);
index <- c(rep("norm",dim(S.normal)[1]),rep("unif",dim(S.unif)[1]));
\end{lstlisting}

The actual model choice is then conducted using the function \texttt{postpr}, which takes as arguments the observed summary statistics $S_{obs}$, the object containing the summary statistics for all models, and the index vector. In addition, the tolerance for the rejection step as well as the method for estimating Bayes factors has to be provided\index{Bayes factors}.  In total, \textit{abc} offers three such methods. The simplest is the the method \texttt{rejection}, which estimates posterior probabilities of the different models directly from the relative proportions of accepted simulations. The two other methods, \texttt{mnlogistic} and \texttt{neuralnet} attempt to correct for the often large tolerance values by estimating the relative densities of retained simulations at $S_{obs}$ using either a multinomial logistic regression  \cite{Beaumont2008,Fagundes2007} or neural networks \cite{Francois2011}, respectively.

The following command will run model choice using the \texttt{neuralnet} method on our toy models and, using the function \texttt{summary()}, print the results to screen in a nice format.

\begin{lstlisting}[language=R]
  model.choice <- postpr(S.obs, index=index, sumstat=allSimulations, tol=0.1, method="neuralnet");
  summary(model.choice);
\end{lstlisting}

The results for our toy models is shown in Table \ref{tab:model_choice}.
As is expected from the observation that the uniform model fails to reproduce the observed summary statistics, the preferred model for this data is the normal model. However, note that the results for the rejection method, which is also run by default when performing a \texttt{mnlogistic} or \texttt{neuralnet} estimation, is much less clear due to the relatively large tolerance applied here.

\begin{table}[!htbp]
\centering
\captionof{table}{Results of model choice on toy models} \label{tab:model_choice}
\begin{tabular}{l c c p{0.1cm} c c}
\toprule
 & \multicolumn{2}{c}{Posterior Probability}& &  \multicolumn{2}{c}{Bayes Factor}\\\cmidrule{2-3}\cmidrule{5-6}
Model & Normal & Uniform & & Normal vs. Uniform\\
\textit{abc} (rejection) & 0.65 & 0.395 & & 1.53\\
\textit{abc} (neuralnet) & 1.00 & 0.00 & & $1.1 \times 10^6$\\
\textit{ABCtoolbox} (GLM) & 1.00 & $9.79\times10^{-13}$ &  & $9.79\times10^{13}$\\
\bottomrule
\end{tabular}
\end{table}

\paragraph{\textit{ABCtoolbox}}\index{ABCtoolbox}
To perform model choice\index{Model choice} with \textit{ABCtoolbox}, simply provide the arguments \texttt{simName} and \texttt{params} for multiple models using semicolons. To run model choice on our toy models, for instance, the input file \texttt{estimate.input} provided abode is modified as follows:

\begin{lstlisting}
task estimate
simName '*\dnormSimulated*';'*\dunifSimulated*'
obsName normal.obs
params 1-2;1-2
maxReadSims 10000
numRetained 1000
maxCor 1.0
\end{lstlisting}

When running \textit{ABCtoolbox} with such an input file, an additional file with tag \texttt{modelFit} is generated. This file contains the marginal densities, Bayes factors and posterior probabilities for each model. The results from this file obtained for our toy models is shown in Table \ref{tab:model_choice}, clearly indicating that the normal model is a much better fit.

\subsection{Model choice validation}\label{modVal}\index{Model choice validation}\index{Validation}
As was shown recently, model choice conducted with ABC may lead to biased or even wrong posterior probabilities, even if the summary statistics are sufficient for all models compared (see \cite{robert_lack_2011}). Consider two models ${\cal M}_1$ and ${\cal M}_2$ of shared parameters $\theta$. If a set of summary statistics $S$ was sufficient for both models, then the likelihood of the summary statistics and the likelihood of the full data are proportional for both models

\begin{align*}
 \P(D|\theta, {\cal M}_1) &= c_1 \P(S|\theta,{\cal M}_1)\\
 \P(D|\theta, {\cal M}_2) &= c_2 \P(S|\theta,{\cal M}_2)
\end{align*}

However, there is no guarantee that the two proportionality constants $c_1$ and $c_2$ are identical, which leads to the Bayes factors\index{Bayes factors} that are off by $c_1/c_2$. Therefore, careful validation is a key and compulsory step of any ABC model choice analysis. An initial first test may be to evaluate the power of choosing the correct model by means of pseudo-observed data sets. Such a cross-validation, which is offered by both \textit{abc} as well as \textit{ABCtoolbox}, simply picks random simulations among those provided from both models, conducts model choice, and records how frequently the correct model was preferred. In addition, \textit{ABCtoolbox} provides means to test for biases in the obtained posterior probabilities by comparing the ABC posterior probability (termed $p_{ABC}$) against those empirically expected ($p_{empirical}$) \citep{Peter2010}.

\paragraph{\textit{abc}}\index{abc (R package)}
The R package \textit{abc} contains the function \texttt{cv4postpr} to conduct cross-validation for model choice. This function randomly picks one simulation from the file containing all simulations, performs model choice using the chosen simulation as pseudo-observed data, and records which model obtained the highest posterior probability. This is then repeated many times to determine the confusion matrix. As an example, consider the following call to \texttt{cv4postpr} to conduct 100 such replicates on our toy models using the index vector created above. To then print the confusion matrix, one may use the function \texttt{summary()} and to obtain a graphical representation the built in \texttt{plot()} function as follows.

\begin{lstlisting}[language=R]
cv.model.choice <- cv4postpr(index,allSimulations, nval=100, tols=0.1, method="neuralnet")
summary(cv.model.choice);
plot(cv.model.choice);
\end{lstlisting}

\paragraph{\textit{ABCtoolbox}}\index{Model choice validation}\index{ABCtoolbox}
To perform model choice validation with ABCtoolbox, simply add the argument \texttt{modelChoiceValidation} followed by the number of pseudo-observed data sets to be used  to the estimation file. To use 1000 pseudo-observed data sets, for instance, you may add the following line to the input file:

\begin{lstlisting}
modelChoiceValidation 1000
\end{lstlisting}

\textit{ABCtoolbox} will then perform cross validation and write the results to two different files. The first has the tag \texttt{confusionMatrix} \index{Confusion matrix}containing the confusion matrix (fraction of correctly and incorrectly inferred models) as well as statistics calculated from it. For the toy model, for instance, we learn from this file that the normal model is correctly identified from data generated under that model in $>99\%$ of the cases.

The second file with tag \texttt{modelChoiceValidation} contains the raw results from the model choice validation and can be used for a more detailed validation analyses. For instance, we have recently proposed to compare the estimated model posterior probabilities with the empirical ones, an analysis that can reveal biases in ABC model choice \citep{Peter2010,Chu2013}. The basic logic of this analysis is that among all pseudo-observed data sets that resulted in an ABC posterior probability $p_{ABC}=x$ in favor of, say, model 1, a fraction $x$ should have been generated under model 1. To test for this, data sets are binned according to their ABC posterior probabilities $p_{ABC}$ and the empirical posterior probabilities $p_{empirical}$ are then estimated as the fraction of simulations within each bin that were indeed generated with model 1.

\begin{figure}[!htbp]
\centering
\includegraphics[keepaspectratio,scale=0.5]{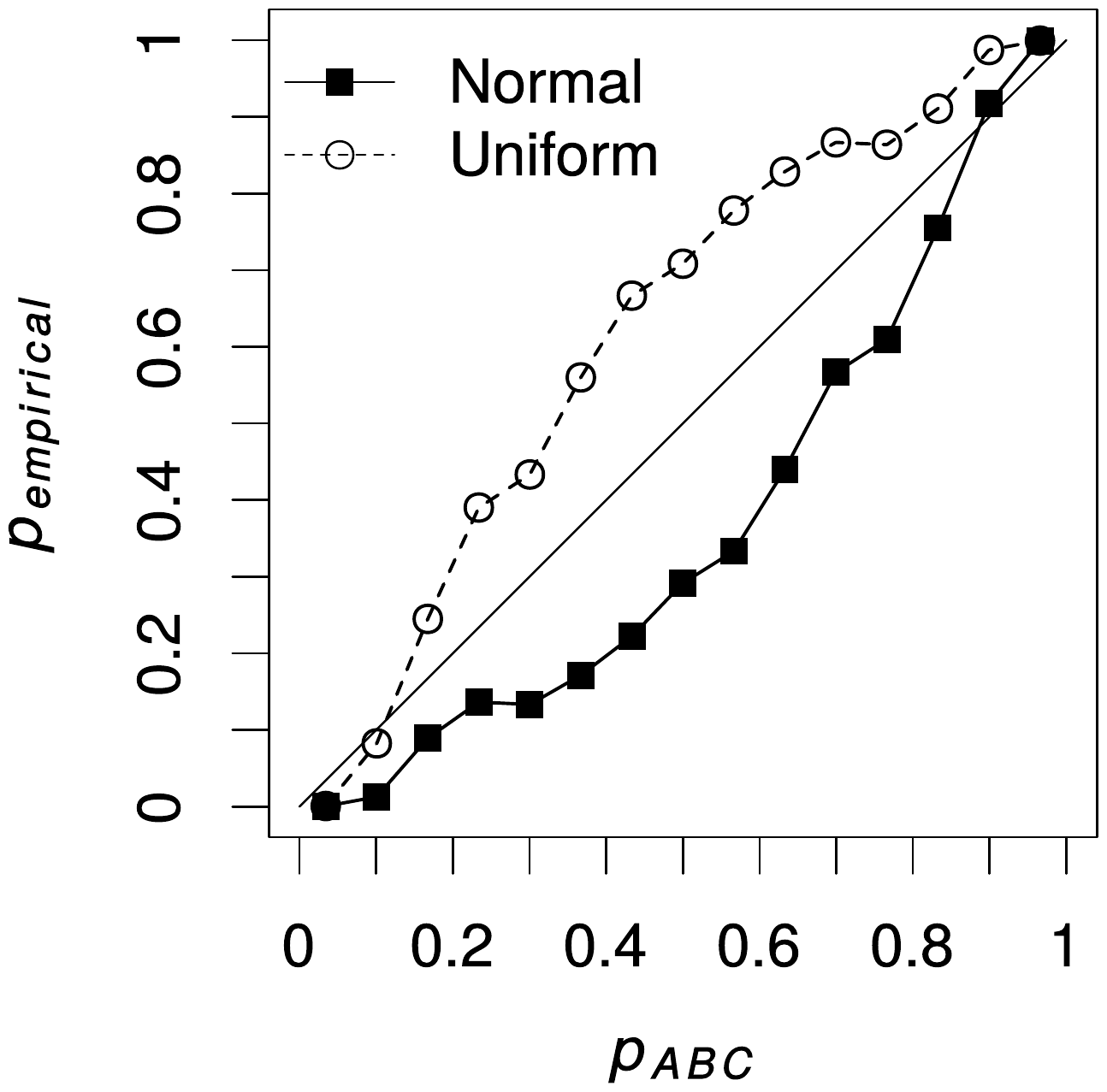}
\caption{\label{Figure_pABC}Posterior probability for normal and uniform distribution models estimated by \textit{ABCtoolbox} ($p_{ABC}$) versus an empirical estimate of the same probability through simulation ($p_{empirical}$).
}
\end{figure}

The \textit{ABCtoolbox} package includes the Rscript \texttt{Make\_Model\_choice\_power\_plot.r} to conduct this analysis and to produce the plot shown in Figure \ref{Figure_pABC}. For our toy models it appears that there is a slight bias towards the normal model. This is evident from the fact that when $p_{ABC}=0.5$, the data sets were actually generated from the uniform model in about $70\%$ of the cases. However, among the data sets that resulted in very high $p_{ABC}$ $(\ge0.99)$, the vast majority was generated under that model. Therefore, we have high confidence in the model choice results of our observed data, which produced  $\ge0.99$ posterior probability support for the normal distribution model (Table \ref{tab:model_choice}).

\subsection{Choosing summary statistics}\label{statsSummary}
\subsubsection{Statistics for parameter inference}\label{statsParamInference}

The choice of summary statistics is crucial in any ABC inference in that too few summary statistics are likely to miss out on important information and too many introducing harmful noise to the estimation \citep{wegmann_efficient_2009,Beaumont2008,blum2013}. To date, many methods have been proposed to choose informative summary statistics from a larger set (see \cite{blum2013} for a review), but we will focus here on those available through the \textit{ABCtoolbox} package, in particular the use of linear combinations of summary statistics.

The use of such linear combinations was first introduced by \cite{wegmann_efficient_2009}, who proposed to find them by means of \index{Partial Least Squares (PLS)} Partial Least Squares (PLS) regression. Broadly speaking, PLS is similar in spirit to a Principal Component Analysis (PCA), but instead of finding linear combinations that maximize the variance explained in the summary statistics space, PLS components are chosen such that they maximize the product of the variance among summary statistics and the covariance between parameters and statistics \citep{Tenenhaus1995}. Recently, alternative means of finding linear combinations of summary statistics have been proposed, such as through boosting \citep{Aeschbacher2012} or by regressing summary statistics on to posterior means inferred from an initial set of simulations \citep{Fearnhead2012}.

\begin{figure}[!htbp]
\centering
\includegraphics[keepaspectratio,scale=0.8]{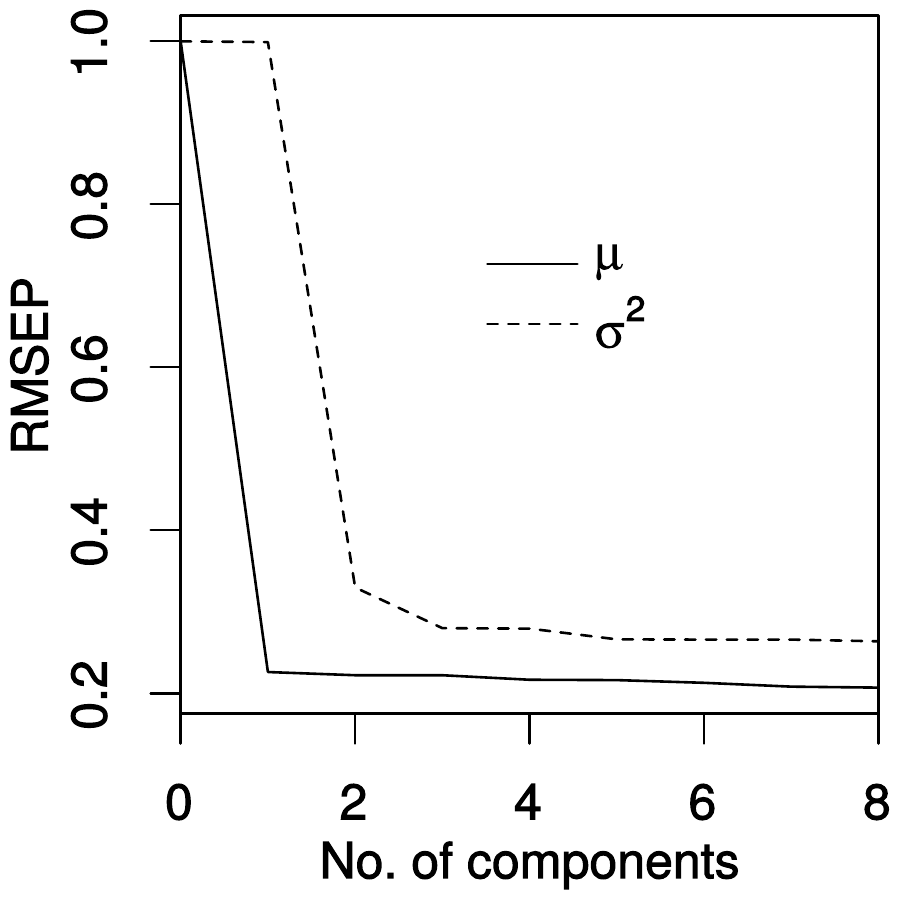}
\caption{\label{Figure_RMSE} Number of PLS components versus root mean square error of prediction (RMSEP) for the normal distribution model.
}
\end{figure}

While all these methods are readily used with \textit{ABCtoolbox} once the linear combinations have been found, we will illustrate the usage of this functionality based on the PLS approach, which is easy to implemevR-package 'pls'. In fact, the \textit{ABCtoolbox} package provides an R-script to perform this analysis taking as input the simulations file (\dnormSimulated). Performing a PLS analysis on the simulations from the normal distribution example reveals that 2 PLS components are sufficient for explaining the variance of the parameters of the normal distribution (Figure \ref{Figure_RMSE}). This result is expected since the mean of a sample and the mean and variance of a sample are sufficient statistics for estimating, respectively, the mean and variance parameters of a normal distribution.

Any definition of linear combinations resulting from such a PLS or any other approach can then be used to transform the statistics of a set of simulations and the observed data using \textit{ABCtoolbox}, and then used in parameter inference. The PLS R-script mentioned above, for instance, writes the resulting PLS components to the file \texttt{PLSdef.txt}, which is then provided to \textit{ABCtoolbox} and run in the \texttt{transform} mode as follows:

\begin{lstlisting}[language=bash]
./ABCtoolbox task=transform linearComb=PLSdef.txt input='*\dnormSimulated*' output=simNorm.pls numLinearComb=2
./ABCtoolbox task=transform linearComb=PLSdef.txt input=normal.obs output=normal.obs.pls numLinearComb=2
\end{lstlisting}

Note that while we provided all arguments to \textit{ABCtoolbox} on the command line, they may equally well be given in an input file. Using the transformed summary statistics in the estimation step is then straightforward: simply provide the transformed files using the arguments \texttt{simName} and \texttt{obsName}.As mentioned above, using alternative ways to find linear combinations is compatible with \textit{ABCtoolbox}, as long as the linear combinations can be written in a definition file as the one created by the PLS script. Alternatively, the statistics may also be transformed with different software and then provided to \textit{ABCtoolbox} (or the R package \textit{abc}) for the estimation step.

One issue with using linear combinations of summary statistics is that information that arises from non-linear combinations of statistics are not taken into account. In such situations, it may be beneficial to increase the summary statistics space through combinations of summary statistics before finding linear combinations \citep{aeschbacher_novel_2012}. As outlined with an example in section \ref{popgenexample}, \textit{ABCtoolbox} has an option (\texttt{doBoosting}) to also generate all pairwise products of summary statistics when generating simulations for this purpose.\index{Boosting}

\subsubsection{Statistics for model choice}
Finding appropriate combinations of statistics for model choice is particularly tricky. Just as for parameter inference, too few statistics may fail to capture important information, while too many are likely adding non-informative noise leading to large estimation variance and potentially a bias. Unfortunately, the methods introduced above for finding good summary statistics for parameter inference are not readily extended to the problem of model choice. If one aims for using linear combinations, the most obvious choice is Linear Discriminant Analysis (LDA), as was recently proposed by \cite{Estoup2012}. To use LDA for model choice with the programs discussed here is similar to the use of linear combinations for parameter inference in that the summary statistics of the observed and simulated data have to be transformed as explained in section \ref{statsParamInference} before running either \textit{ABCtoolbox} or the R package \textit{abc} to perform model choice.

As an alternative to LDA, \textit{ABCtoolbox} offers a \index{Greedy search algorithm for model choice} greedy search algorithm to identify the combination of statistics having the largest power to discriminate between models. This search is done iteratively, by firstly evaluating the power of each single statistic and then adding additional statistics until no increase in power is observed. To perform this type of analysis, \textit{ABCtoolbox} has to be run in the \texttt{findStatsModelChoice} mode and by providing the simulation files for at least two models, as well as the parameters required to perform the estimation. In addition, and using the argument \texttt{modelChoiceValidation}, one also needs to specify the number of simulations to be used as pseudo-observed data in each iteration to evaluate the power. As an example, consider the following input file for performing this type of analysis to find summary statistics appropriate for contrasting the normal versus uniform distribution toy models. We have added argument \texttt{maxCorSSFinder} and set 
it equal to one in order to include combinations of statistics that are highly correlated in the greedy search. A lower threshold might be apporpirate if many summary statistics are used in order to speed up the search. 

\begin{lstlisting}
task findStatsModelChoice
simName '*\dnormSimulated*';'*\dunifSimulated*'
obsName '*\simpleobs*'
maxCor 1.0
maxCorSSFinder 1.0
params 1-2;1-2
numRetained 1000
maxReadSims 10000
outputPrefix ABC_searchStats
modelChoiceValidation 1000
\end{lstlisting}

The results of this analysis are written to a file with tag \texttt{searchStatsgreedySearch}. Part of this file is shown in Table \ref{modelchoicepower}. As shown there, the power to distinguish between these models is very high for multiple sets of summary statistics. Generally, it is recommended to choose the smallest among all sets with highest power, which would be the set consisting of the statistics \texttt{var} and \texttt{range}. As is shown in Figure \ref{Figure_dist},  the two-dimensional distribution of these two statistics is indeed rather different between the models.

\begin{table}[!htbp]
\centering
\captionof{table}{Combinations of statistics sorted by estimated power to distinguish models.} \label{modelchoicepower}
\begin{tabular}{lcccl}
\toprule
Rank&Power&Largest Pairwise Correlation&No.statistics&Statistics\\\cmidrule{1-5}
1&1&0.966&3&mean,var,range\\
6&0.999&0.966&2&var,range\\
82&0.755&0&1&range\\
\bottomrule
\end{tabular}
\end{table}

\section{Generating simulations}\label{autosim}
For simple models such as the normal distribution example that we examined in the previous section, it is relatively easy to perform the simulations using custom scripts written in scripting languages such as R. However, for realistically complex models we often rely on specialized programs for performing simulations. Moreover, for certain ABC variants such as ABC-MCMC, the simulation and estimation procedures are inherently linked, thus requiring running the program that performs simulations jointly with the program that performs ABC. When choosing the appropriate program to do simulations we should keep in mind that interpreted languages, such as R, are generally inefficient. In most cases, compiled languages such as C should be preferred. In this section we will illustrate how to use the program \textit{ABCtoolbox} as well as the R package \textit{EasyABC} to automate and streamline the simulation process and to perform more sophisticated ABC techniques such as ABC-MCMC. 

\subsection{Generating simulations for rejection}
We will first focus on how to use these two ABC pipelines to generate simulations from parameters values drawn from prior distributions. The such generated simulations are then ready to be used with all the estimation techniques introduced above.

\paragraph{\textit{EasyABC}}\label{autosimeasyabc}\index{EasyABC}\index{External simulation software}
This R package allows the user to launch simulations from an external program and to retrieve the output of these simulations in a format ready for post-processing or to dynamically perform ABC-MCMC. To achieve this, the user has to provide both a list containing the definitions of the prior distributions, as well as a model definition. The list containing prior distributions simply contains the names of the desired distributions, along with their arguments. For instance, a list defined as

\begin{lstlisting}[language=R]
prior <- list(c("unif",-1,1),c("unif",0.1,4));
\end{lstlisting}

will imply that there are two model parameters with uniform priors bounded at -1 and 1, and 0.1 and 4, respectively.

The model may be either an R function taking the parameters as arguments and returning a vector of summary statistics, or the name of an executable that will be used to generate the simulation. In case an executable is given, it is assumed that this executable will read the model parameters to be used from a file called \texttt{input} and write the resulting summary statistics to a file called \texttt{output}. These files are read and written dynamically as \textit{EasyABC} concatenates into a list the parameters sampled from the prior and the simulated summary statistics. As an example, consider an executable R-script named \texttt{generate\_norm\_EasyABC.R} that wraps a program to run the simulation of a normal distribution which is a model with two parameters:

\begin{lstlisting}[language=R,title=Standalone R-script to perform simulations from a normal distribution for EasyABC]
#!/usr/bin/Rscript
param<-scan("input")
sampleSize <- 100;
data <- rnorm(sampleSize, mean=param[1], sd=param[2]);

calc.stats <- function (x){
  S <- c(mean(x), var(x), median(x), range(x), max(x)-min(x),
  quantile(x, probs=c(0.25, 0.75)));
  names(S) <- c("mean", "var", "median", "min", "max", "range", "Q1", "Q3")
  return(S);
}

sim <- calc.stats(data);
write.table(t(sim), file="output", quote=F, row.names=F,col.names=F);
\end{lstlisting}

The user should make the script executable like this:

\begin{lstlisting}[language=bash]
chmod +x generate_norm_EasyABC.R
\end{lstlisting}

\textit{EasyABC} is used to generate simulations (in this case $10^3$) with the priors defined above as follows:

\begin{lstlisting}[language=R]
ABC_sim <- ABC_rejection(model=binary_model('./generate_norm_EasyABC.R'),prior=prior,nb_simul=1000)
\end{lstlisting}

which should take approximately 2 minutes to finish. Note that using the internal function of R to generate $10^3$ deviates from a normal distribution would take less than 1 second to complete. Therefore using an external program would be advised only if an R function for performing the simulations cannot be devised (for example for complex population genetics simulations, see below). 

The command above would only generate the simulations. To perform rejection and obtain the posterior distribution of parameters we need to specify the observed summary statistics with argument \texttt{summary\_stat\_target} and the tolerance value with argument \texttt{tol} as follows:

\begin{lstlisting}[language=R]
ABC_sim <- ABC_rejection(model=binary_model('./generate_norm.R'),prior=prior,nb_simul=1000,summary_stat_target=sum_stat_obs,tol=0.1);
\end{lstlisting}

\paragraph{\textit{ABCtoolbox}}\label{autosimABCtoolbox}\index{External simulation software}\index{ABCtoolbox}
An even more advanced and feature-rich way of using existing programs to generate simulations is offered by \textit{ABCtoolbox}. To do so, \textit{ABCtoolbox} has to be run in \texttt{simulate} mode, specified with the argument \texttt{task}. Similarly to \textit{EasyABC}, the user then needs to specify both the model parameters and their prior distributions, as well as how to use existing programs to generate simulations using values drawn from the prior.

The model parameters and their priors have to be provided through an external file referred to as the \texttt{est} file, the name of which is provided with the argument \texttt{estName}. This file is structured in three distinct sections called \texttt{[PARAMETERS]}, \texttt{[RULES]} and \texttt{[COMPLEX PARAMETERS]}. Only the first of those is mandatory and contains the definitions of the model parameters for which estimations are to be carried out. These model parameters and their prior distributions are declared using multiple columns as explained in Table \ref{estNamedeclare}. In brief, the first column indicates whether or not a model parameter is to be truncated to an integer value, the second column lists the name of the parameter and the third column the prior distribution function. The remaining columns contain the parameters for this distribution, for instance the lower and upper bound, as wells as the mean and standard deviation for a normal prior. The last column specifies whether or not the 
parameter values are to be printed to the output file.

\begin{table}[!htbp]
\centering
\captionof{table}{Declaration of parameters in the \texttt{estName} file.} \label{estNamedeclare}
\begin{tabular}{cl}
\hline
Column & Content\\\hline
1 & Indicator 1/0 for being integer or rational number.\\
2 & Name of the parameter.\\
3 & Type of prior (see \textit{ABCtoolbox} Manual for the types of supported priors).\\
4 - & Parameters for prior (for example min,max for uniform prior).\\
Last&Indicator output/hide for whether to print the parameter in the output file.\\
\hline
\end{tabular}
\end{table}

As an example, consider the following \texttt{est} file.
\begin{lstlisting}
[PARAMETERS]
0 PARAM_A unif -1 1 output
0 PARAM_B norm -10 10 1 2 output
[RULES]
PARAM_A > PARAM_B
[COMPLEX PARAMETERS]
0 PARAM_B_SCALED = exp(PARAM_B) / PARAM_A
\end{lstlisting}

Here, we made use of the optional  \texttt{[RULES]} section to limit the simulations to cases where \texttt{PARAM\_A} is larger than \texttt{PARAM\_B}. In addition, we benefited from \texttt{[COMPLEX PARAMETERS]} section to define a new variable \texttt{PARAM\_B\_SCALED}, which will always be set to the exponential of \texttt{PARAM\_B}, divided by the value of \texttt{PARAM\_A}. \textit{ABCtoolbox} will understand most mathematical symbols and offers a wide variety of functions in this section, which allows for the definition of prior distributions and model parameterization in a different way than required by the simulation software.

To demonstrate the use of \textit{ABCtoolbox} to perform simulations we can use a slightly modified simulation script named \texttt{generate\_norm\_ABCtoolbox.R} to generate deviates from a normal distribution similarly to the procedure described above for \textit{EasyABC}:

\begin{lstlisting}[language=R,title=Standalone R-script to perform simulations from a normal distribution for ABCtoolbox]
#!/usr/bin/Rscript
args = commandArgs(trailingOnly=TRUE)
param1=as.numeric(args[1])
param2=as.numeric(args[2])
sampleSize <- 100;
data <- rnorm(sampleSize, mean=param1, sd=param2);

calc.stats <- function (x){
  S <- c(mean(x), var(x), median(x), range(x), max(x)-min(x),
  quantile(x, probs=c(0.25, 0.75)));
  names(S) <- c("mean", "var", "median", "min", "max", "range", "Q1", "Q3")
  return(S);
}

sim <- calc.stats(data);
write.table(t(sim), file="summary_stats-temp.txt", quote=F, row.names=F);
\end{lstlisting}

To specify how \textit{ABCtoolbox} is to interact with the external simulation software, the arguments \texttt{simProgram} and \texttt{simArgs} are used, where the former defines the name of the executable to be used, and the latter the arguments to be passed to that executable. These arguments may contain tags referring to model parameters listed in the \texttt{est} file, as well as any other string. The appropriate input file for \textit{ABCtoolbox} may thus look as follows:

\begin{lstlisting}
task simulate
obsName normal.obs
estName Rules.est
numSims 1000
simProgram generate_norm_ABCtoolbox.R
simArgs PARAM_A PARAM_B
\end{lstlisting}

The \texttt{Rules.est} file that contains the definitions of parameters and their priors for should be specified like this:

\begin{lstlisting}
[PARAMETERS]
0 PARAM_A unif -1 1 output
0 PARAM_B unif 0.1 4 output
\end{lstlisting}

In this example, the parameters are read by the simulation program directly from the command line. In case the simulation program reads the parameters from a specific input file, \textit{ABCtoolbox} can be set up to scan such a file and to replace all occurrences of model parameter tags defined in the \texttt{est} file with their current values, and to save the result to a new file, which is then passed to the simulation program. To make use of this feature, the name of the input file has to be specified with the argument \texttt{simInputName}, and the tag \texttt{SIMINPUTNAME} may then be used to refer to the newly created input file among the arguments passed to the simulation program.

Moreover, the output of the simulation program is stored in a file named "summary\_stats-temp.txt" which is read by \textit{ABCtoolbox} by default but a different name could be specified with argument \texttt{sumStatName}. In case the simulation program is generating data instead of directly summary statistics itself, \textit{ABCtoolbox} can run additional programs to do extra operations on the output of the simulation program, such as the calculation of summary statistics. Such a program can be defined with the argument \texttt{sumStatProgram} and the command-line arguments for the program are set with \texttt{sumStatArgs}. Note that \texttt{sumStatProgram} will always run after \texttt{simProgram}. A list of commonly used arguments when running \textit{ABCtoolbox} in \texttt{simulate} mode are listed in Table \ref{simsettings}.

\begin{table}[!htbp]
\centering
\captionof{table}{\textit{ABCtoolbox} settings for simulation} \label{simsettings}
\begin{tabulary}{1.0\textwidth}{p{2cm}p{3cm}p{8cm}}
\toprule
Setting type & Setting & Description \\\cmidrule{1-3}
\multirow{10}{*}{Basic} & task& Possible options simulate and estimate. \\
&samplerType & Possible sampler types are standard, MCMC, PaSS, PMC. \\
&numSims&No. of simulations to perform. \\
&outName&Prefix for output files. \\
&estName&Filename containing definitions of priors for parameters and rules. \\
&simProgram&Program to perform simulations. \\
&simArgs &Arguments for simulation program. \\
&obsName &File containing observed summary statistics. \\
&sumStatProgram&Program to be run after simProgram. For example a script calculating summary statistics.\\
&sumStatArgs&Arguments for sumStatProgram. \\
&sumStatName &File containing simulated summary statistics. \\
&doBoxCox&Do boxcox transformation.\\
&linearCombName&File containing linear combinations for transformation of statistics. (e.g., PLS components).\\
&doBoosting&Use all product combinations of statistics as additional statistics.\\\cmidrule{1-3}
\multirow{5}{*}{MCMC} &numCaliSims&No. of calibration simulations.\\
&thresholdProp&Tolerance proportion of calibration simulations.\\
&rangeProp&Range of proposals.\\
&startingPoint&Starting location set from a random simulation (random) or the simulation with the minimum distance to the observed data (best).\\
&mcmcSampling&Interval between iterations that are printed in the output file.\\
\bottomrule
\end{tabulary}
\end{table}

\subsection{Performing MCMC}\index{Markov Chain Monte Carlo}\index{ABC-MCMC}
Several other likelihood-free algorithms have been proposed that overcome the inherently low acceptance rates of rejection algorithms, among them a  Markov chain Monte Carlo sampler (ABC-MCMC) first introduced by \cite{marjoram_markov_2003}, a Gibbs sampler using parameter-specific statistics (ABC-PaSS; \cite{kousathanas_likelihood-free_2016}) and sequential Monte Carlo or particle samplers (ABC-PMC; \citep{Sisson2007a, Beaumont2009}). While both the R package \textit{EasyABC} as well as \textit{ABCtoolbox} offer several types of algorithms, we will focus here on the use of ABC-MCMC with these tools.

The basic idea of ABC-MCMC is to replace the likelihood ratio in the Hastings ratio of a classic MCMC by an acceptance-rejection step using some tolerance $\epsilon$. Such an ABC-MCMC chain is then generating samples directly from $\P(\|S - S_{obs})\|<\epsilon|\theta)$, where $\theta$ is the vector of model parameters, $S$ and $S_{obs}$ the simulated and observed vectors of summary statistics, respectively, and $\|\cdot\|$ some distance measure in the summary statistics space. Such an algorithm was shown to require much less simulations than standard ABC methods to obtain equally good posterior estimates \citep{marjoram_markov_2003}. However, it turned out to be relatively tricky to tune this algorithm to perform properly since the acceptance rate of such an algorithm is directly given by the absolute likelihood, rather than the relative likelihood as in standard MCMC. A result of this is that ABC-MCMC chains may easily get stuck in regions of low likelihood, requiring a careful choice of both the tolerance 
$\epsilon$ as well as the initial starting positions. To improve the performance of this algorithm, we have proposed to tune the ABC-MCMC algorithm by means of an initial training set of simulations \citep{wegmann_efficient_2009}, which has been adopted by both \textit{EasyABC} as well as \textit{ABCtoolbox}. Specifically, the idea of such a \index{Calibration (ABC-MCMC)} calibration step is to choose a tolerance value $\epsilon$ that will result in sufficiently high acceptance rates and to find starting values in high likelihood regions. As with the classic rejection algorithm, it may be useful to transform summary statistics when calculating distances \citep{wegmann_efficient_2009}, and hence both \textit{EasyABC} as well as \textit{ABCtoolbox} offer to specify such transformations to be used during an ABC-MCMC chain.

While generally faster, an important issue with ABC-MCMC as well as Sequential Monte Carlo algorithms is that their output can not be directly used for validation. Instead, validation has to be done by repeating the whole process using pseudo-observed data sets, which may easily eat away the computational benefit of using these methods.

\paragraph{\textit{EasyABC}}\label{easyabcmcmc}\index{EasyABC}
In \textit{EasyABC}, the ABC-MCMC algorithm is offered through the function \texttt{ABC\_mcmc()}, which takes similar arguments as the function to perform the rejection algorithm, namely a list with prior definitions as well as a model, but also requires the vector containing the observed summary statistics to be specified using the argument \texttt{summary\_stat\_target}. In addition, several arguments for tuning the actual MCMC run are required. As an example, consider the following R code to generate posterior samples using ABC-MCMC for our normal toy model, using the function \texttt{calc.stats()} and the vector of observed summary statistics \texttt{S.obs} introduced above:

\begin{lstlisting}[language=R]
#define model
toy_model <- function(x){
  data <- rnorm(100, x[1], sqrt(x[2]));
  return(calc.stats(data));
}
toy_prior <- list(c("unif",-1,1),c("unif",0.1,4));

#run ABC-MCMC
ABC_posterior <- ABC_mcmc(method="Wegmann", model=toy_model, prior=toy_prior, n_between_sampling=1,n_rec=10000, summary_stat_target=S.obs, n_calibration=10000, tolerance_quantile=0.1, numcomp=2);
\end{lstlisting}

Here, the argument \texttt{n\_rec} specifies that 10,000 samples are to be generated. Further, the arguments \texttt{n\_calibration} and \texttt{tolerance\_quantile} specify that the ABC-MCMC chain will be calibrated from 10,000 simulations conducted under the prior, of which a fraction of 0.1 will be retained to calibrate the MCMC chain. Finally, the argument \texttt{numcomp} specifies that the total set of summary statistics is to be transformed into 2 PLS components \index{Partial Least Squares (PLS)}.

Since an ABC-MCMC run is generating posterior samples, the output can be used directly to plot posterior distributions.

\begin{lstlisting}[language=R]
par(pty="s",mfrow=c(1,2));
plot(density(ABC_posterior$param[,1],from=-1,to=1,adjust=3),main="",xlab=expression(mu));
plot(density(ABC_posterior$param[,2],from=0.1,to=4,adjust=3),main="",xlab=expression(sigma^2));
\end{lstlisting}

\paragraph{\textit{ABCtoolbox}}\index{ABCtoolbox}
To perform the ABC-MCMC algorithm with \textit{ABCtoolbox}, a few arguments have to be added to the input file shown above for standard sampling. First, the argument \texttt{samplerType} has to be set to \texttt{MCMC}. Then, the arguments \texttt{numCaliSims}, \texttt{thresholdProp} and \texttt{rangeProp} are used to specify the number of simulations to be used for \index{Calibration (ABC-MCMC)} calibration, the fraction of those simulations to be used to calibrate the threshold, and the fraction of the standard deviation of parameter values among these retained simulations to be used to propose new values during the MCMC chain, respectively. To transform the summary statistics during the MCMC chain, a file with the definition of linear combinations can be provided with the argument \texttt{linearCombName}. To use PLS transformations, for instance, an initial set of \index{Calibration (ABC-MCMC)} calibration simulations can be used to find appropriate PLS components as discussed above, and the resulting PLS definition file is then provided using this argument\index{Partial Least Squares (PLS)}. For an 
example of an input file we refer the reader to the population genetics example discussed below.

\subsection{A population genetics example} \label{popgenexample}
Here we will illustrate how to implement techniques described in the previous sections to estimate important aspects of the recent human demographic history from an allele frequency data set made publicly available by \cite{boyko_assessing_2008}. Specifically, we will use the site-frequency spectrum (SFS) for synonymous sites obtained for a sample of 24 African Americans (from Table S2 in \cite{boyko_assessing_2008}) to infer the parameters of a simple population genetic model. The SFS is an information rich summary of allele frequency data and synonymous sites in a gene are sites where any point mutation would lead to the same amino acid, thus likely to evolve neutrally which is an assumption we have to make for demographic inference. Our model assumes an ancestral African population of size $N_{ANC}$ which experienced an instantaneous change in size  $t$ generations ago to $N_{CUR}$. We note that there are multiple full-likelihood solutions available to infer the parameters of this simple model from SFS 
data that might outperform ABC \citep[e.g.][]{Excoffier2013, Gutenkunst2009}. However, the goal here is to provide a detailed step-by-step guide to using  \textit{ABCtoolbox} \index{ABCtoolbox} for demographic inference, for which we prefer a simple model that is fast to run. The benefit 
of ABC over the full likelihood approaches lies in its flexibility, and working through this rather simple example will illustrate all aspects necessary to build even more complex models that may violate the assumptions of available full-likelihood solutions. In Table \ref{required_files} we provide a lookup table of all the files that will be described and used in this example.

\begin{table}[!htbp]
\centering
\footnotesize
\captionof{table}{Files required to run the full population genetics example} \label{required_files}
\begin{tabulary}{1.0\textwidth}{p{3cm}p{6cm}p{6cm}}
\toprule
Filename&Description&Source\\\cmidrule{1-3}
popgen.obs&Observed summary statistics&Section \ref{popgenexample}\\
popgen.est&Rules file for ABCtoolbox containing definition of priors&Section \ref{popgenexample}\\
popgen.input&ABCtoolbox input file&Section \ref{popgenexample}\\
fsc25221&fastsimcoal2 executable&http://cmpg.unibe.ch/software/fastsimcoal2/\\
popgen.par&fastsimcoal2 input file&Chapter appendix\\
calcPopstats.pl&Perl script to calculate summary statistics from fastsimcoal2 output&Chapter appendix\\
findPLS.r&R-script to find PLS components & https://bitbucket.org/phaentu/abctoolbox-public/\\
PLSdef\_popgen.txt&file containing PLS definitions generated with findPLS.r&Chapter appendix\\

\bottomrule
\end{tabulary}
\end{table}

Following \cite{boyko_assessing_2008}, we parameterized the time of the size change in units of the current population size ($\tau=t/(2 \times N_{CUR})$) and to allow the simulations to be performed in a time reasonable for an illustrative example, we downsampled the original data from the original 5 million sites to the SFS of only 10,000 sites shown in Table \ref{popsfstable}. From this 
data, we then calculated the set of classic population genetic summary statistics shown in Table \ref{popstattable}.

\begin{table}[!htbp]
\centering
\footnotesize
\setlength{\tabcolsep}{3pt}
\captionof{table}{Downsampled synonymous SFS.} \label{popsfstable}
\begin{tabular}{lccccccccccccccccccccccccc}
\hline
Site class&0&1&2&3&4&5&6&7&8&9&10&11&12&13&14&15&16&17&18&19&20&21&22&23&24\\\hline
Site count&9906&7&5&2&0&1&1&0&0&1&0&0&0&0&0&0&0&0&0&0&0&0&0&0&77\\
\hline
\end{tabular}
\end{table}

\begin{table}[!htbp]
\centering
\footnotesize
\captionof{table}{Summary statistics for synonymous sites.} \label{popstattable}
\begin{tabular}{lcl}
\toprule
Statistic&Header tag&Value\\\cmidrule{1-3}
No. of singletons&sfs1&7\\
No. of segregating sites ($S$) &S&17\\
Average pairwise diversity ($\pi$)&pi&3.06\\
Waterson's thita&thita&4.55\\
Tajima's $D$&taj\_D&-1.17\\
\bottomrule
\end{tabular}
\end{table}

These summary statistics should then be stored in the file \texttt{popgen.obs} for later usage as follows:
\begin{lstlisting}[title=Observed file popgen.obs]
sfs1 S pi thita taj_D
7 17 3.06 4.55 -1.17
\end{lstlisting}

The first step always consists of defining the model parameters in the \texttt{est} file. For the model concerned here, we will use the file \texttt{popgen.est} provided below.

\begin{lstlisting}[title=Rules file popgen.est]
[PARAMETERS]
0 LOG10_N_CUR    unif 2 6    output
0 LOG10_OMEGA    unif    -3    3    output
0 TAU    unif 0 1    output
0 MUTRATE fixed 2.5e-8 hide
[COMPLEX PARAMETERS]
1    N_CUR = pow10(LOG10_N_CUR)    hide
1    T1 = TAU * 2 * N_CUR    hide
0    OMEGA = pow10(LOG10_OMEGA)    hide
\end{lstlisting}

As can be seen from this file, we decided to put uniform priors on the logarithm of the current population size $N_{CUR}$ and the relative size of the ancestral population $\omega = \frac{N_{ANC}}{N_{CUR}}$, but a uniform prior on the relative age of the population size change $\tau$.

To generate simulations under this model, we will make use of the program \textit{fastsimcoal2} ($v. 2.5.2.21$ downloaded from \href{http://cmpg.unibe.ch/software/fastsimcoal2/}{http://cmpg.unibe.ch/software/fastsimcoal2/}) that allows to simulate SFSs under various demographic scenarios \citep{Excoffier2013} \index{External simulation software}. However, \textit{fastsimcoal2} requires the parameters to be specified differently from our priors, and we thus make use of the \texttt{[COMPLEX PARAMETERS]} section to transform our model parameters appropriately. Specifically, we have to provide $N_{CUR}$ and the population size change on the natural scale, and further the age of the population size changes in generations. We then prepare the input file \texttt{popgen.par} for \textit{fastsimcoal2} that specifies this model, using the parameter tags defined in the input file. While explaining the details of how to use \textit{fastsimcoal2} for such a model is beyond the scope of this chapter, we provide the corresponding input file in the appendix 
and refer the reader to the manual of \textit{fastsimcoal2} for more details. To calculate summary statistics from the simulated data we will use the custom perl script \texttt{calcPopstats.pl} also provided in the appendix to this chapter.

In order to use a \index{Partial Least Squares (PLS)}PLS transformation during the MCMC chain, we first generated an initial set of 1000 simulations using \textit{ABCtoolbox} using the following input file:

\begin{lstlisting}[title=ABCtoolbox input file to perform simulations with fastsimcoal2]
task simulate
obsName popgen.obs
estName popgen.est
numSims 1000
outName popgen_PLS
simInputName popgen.par
simProgram ./fsc25221
simArgs -i popgen-temp.par -s 0 -d -n 1 -q -x
sumStatProgram calcPopstats.pl
sumStatArgs popgen-temp/popgen-temp_DAFpop0.obs
doBoosting
\end{lstlisting}

We specify how \textit{ABCtoolbox} is interacting with \textit{fastsimcoal2} (executable \texttt{fsc25221}) with arguments \texttt{simProgram} and \texttt{simArgs}. We further specify our custom perl script \textit{calcPopstats.pl} with argument \texttt{sumStatProgram} which calculates summary statistics from the output of \textit{fastsimcoal2}. While we refer the reader to the manual of \textit{fastsimcoal2} for the details on the command line used, we note that the output written by \textit{fastsimcoal2} will be located in a subdirectory (\texttt{popgen-temp}) and have a specific name (\texttt{popgen-temp\_DAFpop0.obs}). We thus provide the path to this file to our perl script \texttt{calcPopstats.pl} via command line arguments (using \texttt{sumStatArgs}). In contrast to previously discussed input files we also added the additional argument \index{Boosting} \texttt{doBoosting}, which will tell \textit{ABCtoolbox} to also add all squares and pair-wise products of calculated summary statistics as additional summary 
statistics. This often proves helpful in exploiting non-linear relationships between parameters and statistics when finding linear combinations.

\begin{figure}[!htbp]
\centering
\includegraphics{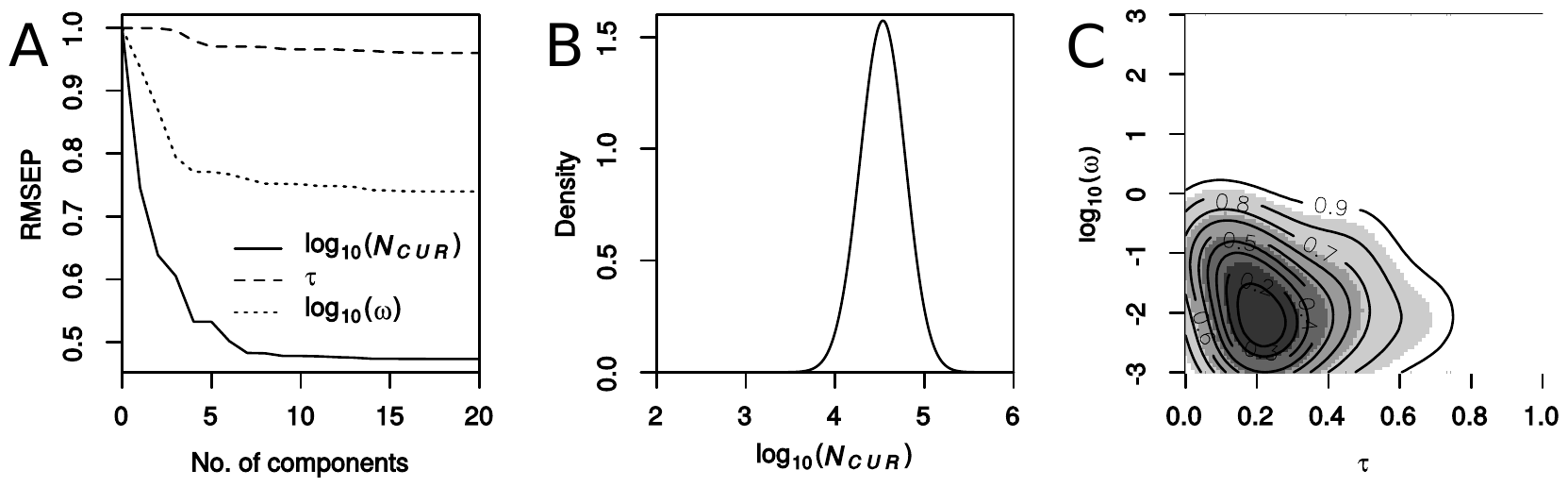}
\caption{\label{Figure_poppost_MCMC} Root mean squared error of prediction (RMSEP) as a function of the number of components used for each parameter (A), marginal posterior estimate for parameter $log_{10}(N_{CUR})$ (B) and joint posterior for $\tau$ and $log_{10}(\omega)$ (C). Solid contour lines specify highest posterior density intervals.
}
\end{figure}

PLS components are then readily identified by following the steps discussed in section \ref{statsParamInference}. By looking at the RMSE (Root-Mean-Squared Error) plot (Figure \ref{Figure_poppost_MCMC}A) we found that 4 PLS components are sufficient to capture the information contained in the total of 20 summary statistics (including the boosted ones). Having the appropriate PLS definition file \texttt{PLSdef\_popgen.txt} at hand, we can then set up \textit{ABCtoolbox} to run an ABC-MCMC chain using the following input file: 

\begin{lstlisting}[title=ABCtoolbox input file to perform ABC-MCMC with fastsimcoal2 and PLS-transformed statistics]
task simulate
samplerType MCMC
obsName popgen.obs
estName popgen.est
numSims 10000
outName popgen_MCMC
simInputName popgen.par
simProgram ./fsc25221
simArgs -i popgen-temp.par -s 0 -d -n 1 -q -x
sumStatProgram calcPopstats.pl
sumStatArgs popgen-temp/popgen-temp_DAFpop0.obs
doBoosting
numCaliSims 1000
thresholdProp 0.1
rangeProp 1
linearCombName PLSdef_popgen.txt
doBoxCox
\end{lstlisting}

This input file differs from the previous one in that the argument \texttt{samplerType} was added to instruct \textit{ABCtoolbox} to run an ABC-MCMC chain, and in that the arguments required for the \index{Calibration (ABC-MCMC)} calibration step (\texttt{numCaliSims}, \texttt{thresholdProp} and \texttt{rangeProp}), and those to use linear combinations of summary statistics (\texttt{linearCombName} and \texttt{doBoxCox}) were added. Note that the R-script \textit{findPLS.r} to find linear combinations provided by \textit{ABCtoolbox} performs a Box-Cox transformation on the summary statistics, and hence in order to use the generated PLS definition file, \textit{ABCtoolbox} needs to perform a similar transformation first, which is requested with the argument \texttt{doBoxCox}. If the user does not wish to perform a PLS transformation and simply use the raw statistics for inference then they can omit arguments \texttt{linearCombName} and \texttt{doBoxCox} from the input file.

While the output of the ABC-MCMC run (\texttt{popgen\_MCMC\_sampling1.txt}) already corresponds to samples taken from the posterior distribution $\P(\|S - S_{obs}\|<\epsilon|N_{CUR}, \omega, \tau)$, an additional improvement may be achieved by conducting an ABC-GLM estimation with an additional rejection step that will further reduce the threshold $\epsilon$ and hence the accuracy of the posterior. Such an analysis can be conducted as described in Section \ref{ABCGLM} and the resulting posteriors may then be plotted in R.

The posterior estimates we obtained for parameters $N_{CUR}$, $\omega$ and $\tau$ (Figure \ref{Figure_poppost_MCMC} B and C) indicated a large population expansion that happened $\sim13000$ generations ago to a present effective population size of $\sim32000$. The credible intervals of the posterior estimates for $\omega$ and especially $\tau$ are large (Figure \ref{Figure_poppost_MCMC}C) due to the small size of the downsampled data set. Additionally, the statistics used here seem not to be informative for parameter $\tau$ as indicated by the PLS analysis in Figure \ref{Figure_poppost_MCMC}A. However, the more precise estimates for $N_{CUR}$ and $\omega$ are in good agreement with the findings of \cite{boyko_assessing_2008}, who used a maximum likelihood method on the full data.

\newpage
\bibliographystyle{plainnat}
\bibliography{bibliographyABC}

\newpage

\section{Appendix}
Here we provide additional files required to replicate our population genetics example. First, we provide the input file \texttt{popgen.par} for \textit{fastsimcoal2} specifying the population genetics model used. Note that we decided to simulate the SFS using 10 independent loci with 1000 sites each.

\begin{lstlisting}[title=\textit{fastsimcoal} input file]
//Number of population samples (demes)
1
//Population effective sizes (number of genes)
N_CUR
//Sample sizes
24
//Growth rates negative growth implies population expansion
0
//Number of migration matrices : 0 implies no migration between demes
0
//historical event: time, source, sink, migrants, new size, new growth rate,migr.matrix
1 historical events
T1 0 0 1 OMEGA 0 0
//Number of independent loci [chromosome]
10 0
//Per chromosome: Number of linkage blocks
1
//per Block: data type, num loci, rec. rate and mut rate
DNA 1000 0.00000 MUTRATE 0.33
\end{lstlisting}

\clearpage
Further, we provide the custom perl script \texttt{calcPopstats.pl} used to calculate summary statistics from site frequency spectra simulated with the program \textit{fastsimcoal2}.

\begin{lstlisting}[title=Perl script to calculate statistics from SFS,language=Perl]
#!/usr/bin/perl -w
use strict;
#read fastsimcoal output SFS file
my $sfsfile=$ARGV[0];
open(FILE,"<",$sfsfile) or die "can't open SFS file";
open (OUT, ">","summary_stats-temp.txt") or die "can't open sum-stats file";
my ($firstline,$header,$sfsline)=(<FILE>,<FILE>,<FILE>);
#split sfsline into sfs
my @SFS=split /\t/,$sfsline;
my @stats;
#calculate stats
my ($sum,$S,$a1,$a2,$taj_D)=(0,0,0,0,0);
my $n=@SFS-2;
my ($b1,$b2)=(($n+1)/(3*($n-1)),2*($n**2+$n+3)/(9*$n*($n-1)));
#No. Segregating. sites S
for (my $i=1;$i<$n;$i++){
$sum=$sum+$i*($n-$i)*$SFS[$i];
$S=$S+$SFS[$i];
($a1,$a2)=($a1+1/$i,$a2+1/$i**2);
}
#Thita and pi
my ($thita,$pi)=($S/$a1,2*$sum/($n*($n-1)));
#Tajima's D
my ($c1,$c2)=($b1-1/$a1,$b2-($n+2)/($a1*$n)+$a2/($a1**2));
my ($e1,$e2)=($c1/$a1,$c2/($a1**2+$a2));
if($S>0) {$taj_D=($pi-$S/$a1)/sqrt($e1*$S+$e2*$S*($S-1));}
#print out stats
@stats=($SFS[1],$S,$pi,$thita,$taj_D);
print OUT join("\t","sfs1","S","pi","thita","taj_D"),"\n",join("\t",@stats,"\n");
close(FILE);close(OUT);system("rm $sfsfile");

\end{lstlisting}

Finally, we provide the file \texttt{PLSdef\_popgen.txt} specifying the PLS transformation of the 20 statistics (7 polymorphism statistics + their products) to 4 components. The first six columns in the file specify the boxcox transformation of the statistic and the remaining 4 columns specify the PLS components.

\begin{lstlisting}[title=The file PLSdef\_popgen.txt containing the definitions of the PLS transformations for ABCtoolbox]
sfs1 1140 0 -17.58 1.06 0.06 0.07 0.22 0.27 -0.16 0.23
S 4478 0 -10.3 1.12 0.12 0.13 0.25 0.1 -0.3 0.11
pi 1730.18 0 -11.52 1.1 0.09 0.11 0.26 0.05 -0.27 0.07
thita 1199.16 0 -10.3 1.12 0.12 0.13 0.25 0.1 -0.3 0.11
taj_D 2.87 -2.19 0.61 1.48 0.51 0.19 0.11 -0.38 -0.53 -0.5
sfs1_X_sfs1 1299600 0 -20 1.02 0.01 0.02 0.21 0.29 0.13 -0.14
sfs1_X_S 5104920 0 -20 1.03 0.01 0.03 0.25 0.16 0.18 -0.12
sfs1_X_pi 1426520 0 -20 1.03 0.02 0.03 0.25 0.12 0.19 -0.11
sfs1_X_thita 1367040 0 -20 1.03 0.01 0.03 0.25 0.16 0.18 -0.12
sfs1_X_taj_D 634.38 -559.99 -0.61 1.48 0.66 0.09 0.15 -0.36 0.45 -0.01
S_X_S 20052500 0 -18.79 1.06 0.04 0.06 0.26 0.03 0.06 -0.08
S_X_pi 7455340 0 -20 1.06 0.03 0.05 0.26 -0.02 0.08 -0.07
S_X_thita 5369820 0 -18.79 1.06 0.04 0.06 0.26 0.03 0.06 -0.08
S_X_taj_D 8737.33 -1796.63 -9.09 1.21 0.61 0.06 0.18 -0.4 0.16 0.14
pi_X_pi 2993510 0 -20 1.05 0.02 0.04 0.26 -0.07 0.12 -0.05
pi_X_thita 1996450 0 -20 1.06 0.03 0.05 0.26 -0.02 0.08 -0.07
pi_X_taj_D 3508.27 -349.33 -13.94 1.14 0.35 0.05 0.18 -0.39 0.07 0.16
thita_X_thita 1437980 0 -18.79 1.06 0.04 0.06 0.26 0.03 0.06 -0.08
thita_X_taj_D 2339.76 -481.12 -9.09 1.21 0.61 0.06 0.18 -0.4 0.16 0.14
taj_D_X_taj_D 8.21 0 -6.67 1.12 0.13 0.12 0.06 -0.25 -0.3 0.78
\end{lstlisting}

\printindex
\end{document}